\documentclass{aa}
\usepackage{amsmath}
\usepackage{graphicx}
\usepackage[scriptsize]{subfigure}
\usepackage{lipsum}
\usepackage{natbib}

\title{RESOLVE: A new algorithm for aperture synthesis imaging \\ of extended emission in radio astronomy}

\author{H. Junklewitz\inst{\ref{inst1},\ref{inst2}} \and M. R. Bell\inst{\ref{inst1}} \and M. Selig\inst{\ref{inst1},\ref{inst2}} \and T. A. En\ss lin\inst{\ref{inst1},\ref{inst2}}} 

\institute{Max-Planck Institut f\"ur Astrophysik (Karl-Schwarzschild-Str. 1, D-85748 Garching, Germany)\label{inst1} \and Ludwig-Maximilians-Universit\"at M\"unchen (Geschwister-Scholl-Platz 1, D-80539  M\"unchen, Germany)\label{inst2}}

\abstract{We present \textsc{resolve}, a new algorithm for radio aperture synthesis imaging of extended and diffuse emission in total intensity. The algorithm is derived using Bayesian statistical inference techniques, estimating the surface brightness in the sky assuming a priori log-normal statistics. \textsc{resolve} not only estimates the measured sky brightness in total intensity, but also its spatial correlation structure, which is used to guide the algorithm to an optimal reconstruction of extended and diffuse sources. For a radio interferometer, it succeeds in deconvolving the effects of the instrumental point spread function during this process. Additionally, \textsc{resolve} provides a map with an uncertainty estimate of the reconstructed surface brightness. Furthermore, with \textsc{resolve} we introduce a new, optimal visibility weighting scheme that can be viewed as an extension to robust weighting. In tests using simulated observations, the algorithm shows improved performance against two standard imaging approaches for extended sources, Multiscale-CLEAN and the Maximum Entropy Method.}


\begin{document}

\titlerunning{RESOLVE: A new algorithm for aperture synthesis imaging in radio astronomy}

\maketitle

\section{Introduction}

Aperture synthesis techniques using large interferometers have a long and successful history in radio astronomy \citep{Ryle, 1986isra.book.....T, RadioSaga}. While enabling observers to achieve very high resolutions, data processing is considerably more complicated than with a single dish instrument. A radio interferometer effectively measures the Fourier transformation of the sky brightness \citep[see e.g.][]{1986isra.book.....T}. Unfortunately, inverting this relationship to achieve an estimate of the desired source brightness is a non-trivial task since an interferometer only samples a fraction of the Fourier plane, effectively convolving the true image brightness with an observation-dependent point-spread function. A crucial part in data reduction is therefore the \textit{imaging}, i.e.~estimating the sky brightness distribution from the observed data. 

To date the most successful and widely used imaging algorithm in radio astronomy is CLEAN \citep{1974A&AS...15..417H}. It assumes the image to be comprised of uncorrelated point sources and iteratively approximates the true image with a large set of delta functions. CLEAN has been demonstrated to be very accurate for observations of point source dominated fields \citep{1986isra.book.....T,WhiteBook,ImagingReview} and over time many variants and more elaborate extensions have been developed to improve various aspects of its performance \citep{ClarkClean,CSCLean,MSClean,MFClean,MSMFClean}. 

However, there are drawbacks with the CLEAN algorithm. Since it effectively  assumes the image to be a large superposition of point sources, its performance is naturally non-optimal for highly resolved, extended and diffuse sources \citep{ImagingReview}. Some of the newest enhancements of CLEAN try to address this problem using a multiscale approach, assuming differently scaled kernel functions like Gaussians instead of sharp delta peaks \citep{MSClean,MSMFClean}, but it is still not clear how to properly choose the scales. Another important drawback of CLEAN is that it is not known how to appropriately propagate measurement uncertainty \citep[e.g.][]{1986isra.book.....T,WhiteBook} and thus, no reliable uncertainty estimates are available.
 
There are other approaches than CLEAN that try to address the problem of imaging extended sources. Among them are the Maximum Entropy Method (MEM) \citep{MEM}, the non-negative-least-squares (NNLS) approach, which has been shown to improve over CLEAN on mildly extended sources \citep{BriggsPaper,ImagingReview}, and, approaches using wavelets within the framework of \textit{Compressed Sensing} \citep{WiauxScaifePaper,SARA,PURIFY}. We will come back to these in Sec.~\ref{SubSec: CLEANaMEM}. 

In this paper, we introduce \textsc{resolve} (\textbf{R}adio \textbf{E}xtended \textbf{SO}urces \textbf{L}ognormal decon\textbf{V}olution \textbf{E}stimator), a novel algorithm for the imaging of diffuse and extended radio sources in total intensity. A new approach to the problem is taken, using Bayesian statistics in the framework of \textit{Information Field Theory} \citep{IFT} and based on clearly formulated mathematical principles. \textsc{resolve} is designed to fulfill two main requirements: 

\begin{enumerate}
 \item It should be optimal for extended and diffuse radio sources. 
 \item It should include reliable uncertainty propagation and provide an error estimate together with an image reconstruction.
\end{enumerate}

An important incentive for the development of \textsc{resolve} are recent advances in radio astronomical instrumentation. The new generation of radio telescopes, such as the upgraded VLA, LOFAR, the SKA pathfinder missions or ultimately the SKA itself, are opening new horizons in radio astronomy \citep[see e.g.][]{TelescopeReview}. Their unprecedented capabilities of simultaneous, broadband frequency coverage including previously unexplored wavelength regimes, sensitivity, and wide fields of view, while still being sensitive to a large range of spatial frequencies, will almost certainly advance astrophysical and cosmological sciences \citep[see e.g. the German SKA white paper,][]{FutureScienceReview}. At the same time, new developments in signal processing and data analysis will be required to exploit these new capabilities. For instance, as yet unreached levels of sensitivity allow in principle for more detailed detection of structures in diffuse emission. \textsc{resolve} takes advantage of this and uses the rich correlation structure prominently present in such high sensitivity data to guide itself toward an optimal reconstruction of extended sources. 

The main astrophysical focus of \textsc{resolve} is by definition on extended and diffuse radio sources. Among those are galaxy clusters with their weak diffuse halos and strong extended relic structures, lobes of radio galaxies, giant radio galaxies, supernova remnants, galactic radio halos, and the radio emission from the Milky Way.

Ultimately, with this paper, we do not only aim to present a new algorithm but we also propose and discuss a statistical framework (see Sec.~\ref{SecTwo}) that, we believe, will be advantageous to formulate and solve upcoming and more complex imaging problems in radio data analysis. Among these could be for instance multi-frequency techniques for GHz - broadband data, direction-dependent calibration problems, unknown beam reconstructions, polarization imaging, and many more. We will come back to an outlook in Sec. \ref{Sec: Conclusions}.

\section{The algorithm}\label{SecTwo}

\subsection{Aperture Synthesis}

In aperture synthesis, we try to connect an array of telescopes in such a way that we can effectively synthesize a combined instrument with a much larger aperture and therefore resolution. Using the van Zittert-Cernike theorem from the theory of optical coherence \citep{Born}, it can be shown that such a radio interferometer takes incomplete samples of the Fourier transformed brightness distribution in the sky \citep{1986isra.book.....T}. We would like to measure a signal, the sky brightness distribution $I$, which is a real, continuous function of position in the sky. In the most basic model, taking an observation of the signal $I$ translates into 

\begin{align}
V(u,v,w) = \ &W(u,v, w) \int \mathrm{d}l \ \mathrm{d}m \frac{I(l,m)}{\sqrt{1-l^2-m^2}} \notag\\
	 & \mathrm{e}^{-2 \pi i \left(ul + vm + w\sqrt{1-l^2-m^2}\right)}. \label{basicequationfull}
\end{align}

The quantity $V(u,v,w)$ is the \textit{visibility} function following classical terminology of optical interferometry. The coordinates $u$, $v$, and $w$ are vector components describing the distance between a pair of antennas in an interferometric array, where this distance is usually referred to as a \textit{baseline}. They are given in numbers of wavelengths, with $u$ and $v$ usually parallel to geographic east-west and north-south, respectively, and $w$ pointing in the direction of the center of the image plane (i.e. the phase center). The coordinates $l$ and $m$ are a measure of the angular distance from the phase center along axes parallel to $u$ and $v$, respectively. $W(u,v,w)$ is a sampling function defined by the layout of the interferometric array. It is zero throughout most of the $u,v,w$-space, apart from where measurements have been made where it is taken to be unity.

For simplicity, we now restrict ourselves to the common approximation of measuring the sky as flat in a plane tangent to the phase center of the observation, such that $w\sqrt{1-l^2-m^2} \approx 0$. Nevertheless, we note that this is not a necessary requirement of our formalism (see Sec.~\ref{SIRA}).  

With this assumption, (\ref{basicequationfull}) simplifies approximatively to a two-dimensional Fourier transformation

\begin{align}
V(u,v) &\approx W(u,v) \int \mathrm{d}l \ \mathrm{d}m \ I(l,m) \ \mathrm{e}^{-2 \pi i \left(ul + vm\right) }. \label{basicequation}
\end{align}

The visibility function is what our instrument measures, but we are actually interested in the brightness distribution of the source in the sky. This means that we ideally want to invert the relationship (\ref{basicequation}). Unfortunately, this is not possible, since we have lost all information on the Fourier modes that have not been measured due to the incomplete sampling of the Fourier plane. Thus, an inversion of (\ref{basicequation}) gives us not the true brightness distribution, but its convolution with the inverse Fourier transform of the sampling function, better known as the \textit{point spread function} (psf) or, in common radio astronomical terminology, the \textit{dirty beam} $I_{db} = \mathcal{F}^{-1}W$:

\begin{equation}
I_{\mathrm{D}} = \mathcal{F}^{-1}V =  \mathcal{F}^{-1}W \mathcal{F}I = I_{db} \ast I \label{invprob}.
\end{equation}
Here, we have introduced a symbolic Fourier operator $\mathcal{F}$ to be strictly defined later, the common notation $I_{\mathrm{D}}$, \textit{dirty image}, for the simple Fourier inversion of the visibilities, and the symbol $\ast$ to denote a convolution operation.

Reconstructing the real brightness distribution is therefore an ill-posed inverse problem. In principle, infinitely many signal realizations could have led to the measured visibility function and we have no way to \textit{exactly} discriminate between them. However, we can find a statistical description that may produce the \textit{most probable} signal given the measured visibility function. 

\subsection{Signal Inference in Radio Astronomy} \label{SIRA}

In the following, we develop a statistical solution to the inverse problem (\ref{basicequation}) using Bayesian inference techniques. Later, under the condition of a spatially extended source brightness distribution, this will lead us to the formulation of \textsc{resolve}. Our derivation relies on notation and methods developed within the framework of \textit{information field theory} \citep{IFT, IFT2}. 

To start, we condense our mathematical notation considerably by rewriting all equations using indexed quantities. If we properly define the Fourier operator in (\ref{invprob}) as $\mathcal{F}_{kx} = \exp(-i(ul + vm))$ with $x=(l,m)$ and $k=(u,v)$, (\ref{basicequation}) becomes

\begin{align}
V_{k} & = W_{k} \int dx \ \mathcal{F}_{kx}I_{x} \notag \\
      & = W \mathcal{F}I. \label{basicequation condensed}
\end{align}

We have now translated our functions and operations on them into a notation that allows us to interpret them as vectors and operators defined on an arbitrary-dimensional functional vector space. For the sake of brevity, we will often even drop the indices and use a notation as in the second line of (\ref{basicequation condensed}). We can do that if we define the inner product between vectors and operators appropriately for discrete and continuous spaces:

\begin{align}
&\mathrm{discrete \ space:} \hspace{10pt} && a^{\dagger}b := \sum_{x} \ \mathcal{V}_{x} \ \overline{a_{x}} \ b_{x} \notag \\
&\mathrm{continuous \ space:} \hspace{10pt} && a^{\dagger}b := \int dx \ \overline{a(x)} \ b(x) \ dx 
\end{align}

where the $\dagger$ symbol stands for a transposing operation (and a possible complex conjugation in case of a complex field). In contrast, where needed explicitly, the $\cdot$ symbol will denote component-wise multiplication, so that $(a \cdot b)_x = a(x) \ b(x)$. The symbol $\mathcal{V}_{x}$ indicates the possible need for a volume factor in the sum, if the inner product actually is just a discretized version of a continuous one. In practice, this is unavoidable, since all quantities effectively become discrete when finally calculated on a computer \citep[for details see][]{nifty}.

That way, we now can effortlessly combine discrete and continuous quantities in our notation. This is important, since, in real observations, the visibility $V_{k}$ is always a function defined over a discrete, complex Fourier space, spanned by $n_d$ measurements, whereas the sky brightness $I_{x}$ is in principle a continuous function, defined over an infinitely large, real space.  

Following the notation of \citet{IFT}, we define two fundamental quantities, the signal $s$ and the data $d$. The signal is the ideal, true physical quantity we would like to investigate with our observation. The data is what our measurement device has delivered us. In this radio astronomical application, the signal is the true brightness distribution in the sky $s := I(l,m)$ and the data is our visibility function $d := V(u,v)$ including measurement noise. From now on, we will use this definition, but will occasionally translate equations into traditional radio astronomical notation for a more transparent presentation.

If we know how to translate the actions of our measurement device into mathematical operations, we can write down a fundamental data model, connecting signal $s$ and data $d$ with a response operator $R$

\begin{equation}
d = Rs. 
\end{equation}
ignoring measurement noise for a moment.

This is basically equation (\ref{basicequation condensed}), if we identify the response operator with

\begin{equation}
R = W \mathcal{F},  
\end{equation}

We can add more terms to this response operator, slowly introducing more complexity. An inevitable addition is to consider a gridding and degridding operation within the sampling $W'= WG$. This is not a feature of the instrument itself, but is needed in its computational representation for purely numerical reasons to put the visibilities onto a regularly spaced grid, in order to apply the Fast Fourier Transform algorithm \citep{CooleyTukey,Bracewell}, improving computational speed enormously:

\begin{equation}
R = W'\mathcal{F} \label{gridresponse}
\end{equation}
Henceforth, if not explicitly shown, we drop the prime and consider $G$ to be contained in the sampling operator $W$.
 
An important extension might be to introduce a mathematical representation of the antenna sensitivity pattern on the sky, usually called primary beam $A$:

\begin{equation}
R = W \mathcal{F} A. \label{response}
\end{equation}

Even more sophisticated instrumental effects like beam smearing or directional dependent sampling could as well be included here. Also an extension of the response to non-coplanar baselines, and thus allowing for a non-negligible $w$ - term in Eq.~(\ref{basicequationfull}), could be directly incorporated without fundamental complication, e.g. in similar form to the $w$ - projection algorithm \citep{WProjection}.  

Another relevant extension is to include multi-frequency synthesis by adding a new dimension to signal and data using e.g. a common spectral model $I(x,\nu) = I(x,\nu_{0}) \left(\frac{\nu}{\nu_{0}}\right)^{-\alpha(x)}$:

\begin{align}\label{Rmfs}
V_{k'} & = \int dx \ R_{kx} I_{x\nu} \notag \\
	& = W_{k} \int dx \ \mathcal{F}_{kx} \ A_{x} \ I_{x\nu_{0}} \left(\frac{\nu}{\nu_{0}}\right)^{-\alpha_{x}} 
\end{align}
with $k' = k\nu$. 

Going a step further, a full approach using all four Stokes polarizations is conceivable. In that case, the response representation can in principle be expanded into a full RIME (radio interferometer measurement equation) description, as presented e.g. by \citet{SmirnovI,SmirnovII}.

However, both, multi-frequency and polarization imaging, are outside the scope of the present work.

In a real observation, our data is always corrupted by measurement noise. This means we have to add such a noise contribution $n$ to our data model:

\begin{equation}
d = Rs + n. \label{invprob2}
\end{equation}

As already noted, even without noise, we cannot exactly invert this relationship. We thus instead seek for the optimal statistical solution for the signal $s$ given our data $d$. To find the optimal reconstruction, we regard the signal as a random field following certain statistics and being constrained by the data. In probabilistic terms, we look for an expression of the \textit{posterior distribution} $\mathcal{P}(s|d)$ of the signal $s$ given the data $d$. It expresses how the data constrain the space of possible signal realizations by quantifying probabilities for each of them. It comprises all the information we might have obtained through a measurement. 

With the posterior probability, we can in principle estimate the real signal by calculating for instance its posterior mean $\left<s\right>_{\mathcal{P}(s|d)}$, equivalent to minimizing the posterior-averaged $\mathcal{L}_2$ - norm of the quadratic reconstruction error $\textrm{argmin}_m \left<\Vert(s-m)\Vert_{\mathcal{L}_2}\right>_{P(s|d)}$ \citep[see e.g.][]{IFT}. This is exactly the type of solution to the ill-posed inverse problem (\ref{invprob2}) that we want. 

Probability theory shows that we can calculate $\mathcal{P}(s|d)$ if we have expressions for the \textit{likelihood distribution} $\mathcal{P}(d|s)$, describing our model of the measurement process and the noise statistics, and for the statistics of the signal alone, the \textit{prior distribution} $\mathcal{P}(s)$. The renowned Bayes' theorem states this as 

\begin{equation}
\mathcal{P}(s|d) =  \frac{\mathcal{P}(d|s) \mathcal{P}(s)}{\mathcal{P}(d)}
\end{equation}

where $\mathcal{P}(d)$ is called the \textit{evidence distribution}. It effectively acts as a normalization factor since it does not depend on $s$ and thus is unimportant for statistical inferences on the signal. 

To specify the likelihood for a radio interferometer observation, we only need a good model for the measurement process. With (\ref{invprob2}), we see that this involves detailed knowledge of the instrument response $R$ and the statistical properties of the measurement noise $n$. 

Throughout this work we will assume the response representation (\ref{response}) to be exact, or expressed differently, the data to be fully calibrated. On the perspective of combining calibration and imaging into one inference step see Sec.~\ref{Sec: Conclusions}. 

As for the thermal noise of a radio interferometer, it is fair to assume Gaussian statistics, mainly induced by the antenna electronics and independent between measurements at different time steps of the observation \citep{1986isra.book.....T}. Henceforth, the noise field $n$ will be assumed to be drawn from a multivariate, zero mean Gaussian distribution of dimension $n_d$:

\begin{align}
\mathcal{P}(n) &= \mathcal{G}(n,N) \notag\\
		 &:= \frac{1}{\mathrm{det}(2 \pi N)^{1/2}} \ \exp\left(-\frac{1}{2} n^{\dagger} N^{-1} n\right). 
\end{align}

The assumption of uncorrelated Gaussian noise leads to a diagonal covariance matrix $N_{kk'} = \delta_{kk'} \sigma_{k}^2$. For this work, we will assume the noise variance $\sigma_{k}^2$ to be known.

We can now derive an expression for the likelihood by marginalizing over the noise field:

\begin{align}
\mathcal{P}(d|s) & = \int \mathcal{D}n \ \mathcal{P}(d|s,n) \ \mathcal{P}(n) \notag\\
		 & = \int \mathcal{D}n \ \mathcal{P}(d|s,n) \ \mathcal{G}(n,N) \notag\\
                 & = \int \mathcal{D}n \ \delta(n - (d - Rs)) \ \mathcal{G}(n,N) \label{likeli1} \\ 
		 & = \mathcal{G}(d-Rs,N) \label{likeli2},
\end{align}
 
where the integral is meant to be taken over the infinite space of all possible noise realizations. By inserting the delta function in (\ref{likeli1}) we have stated the implicit assumption that our response (\ref{response}) is exact.

We are left with the crucial question of how to statistically represent our signal. Until now, the derivation was kept general and we effectively formulated an inference framework for aperture synthesis imaging. Now, we need to specify a prior $\mathcal{P}(s)$, depending on the type of signal field to which the statistical estimation should be optimal.

In the next section, we present a solution to the inference problem with a signal prior chosen to represent the properties of extended and diffuse emission.

\subsection{\textsc{RESOLVE}: Radio Extended Sources Lognormal Deconvolution Estimator}\label{Sec:MainAlgo}

To specify the prior distribution, we choose to follow an approach of least information. The question is: What is the most fundamental, minimal state of knowledge we have about the signal, prior to the measurement and without introducing any specific biases? 

In this work, we want to focus on diffuse and extended sources in total intensity. Stating this alone enables us to give a few central assumptions we want to be reflected in the prior distribution:

\begin{enumerate}
 \item An extended source exhibits a certain, \textit{a priori} translationally and rotationally invariant (but usually unknown) spatial correlation structure.
 \item The signal field must be strictly positive, since it should represent a physical intensity.
 \item Typically, signal fields in radio astronomy show high variation in structures across the observed field of view, with a few strong components surrounded by weak extended structure, going over to large regions basically dominated by noise, usually spanning many orders of magnitude in intensity. \label{basicenum}
\end{enumerate}

Apart from these statements, we assume that we know nothing more specific about our signal, and the prior should be chosen accordingly. For instance, we do not want to include specific source shapes or intensity profiles. 

The assumption of translational and rotational invariance is very common and useful in signal inference, where it translates into homogeneity and isotropy of the prior statistics. Given our just stated, restricted prior assumptions, there is no reason, in general, to assume a priori that the correlation of the signal should change under spatial translation or rotation\footnote{It should be emphasized that this \textit{a priori} assumption is not in contradiction with an \textit{a posteriori} solution not exhibiting homogeneity and isotropy. Ultimately, if the combination of data and measurement noise allow for a specific source shape, the likelihood will dominate the prior and drive the reconstruction in this direction.}. We thus keep this assumption as valid throughout this paper.

The first constraint (1.) urges us to consider how to include the fact that the signal exhibits a spatial correlation of unknown structure. First we might argue just to use an uninformative prior, not favoring any particular configuration. But, in fact, we do know something, namely that there \textit{is} a spatial correlation, although its exact structure is obscure to us. Thus, we search for the statistics of a random field about whose correlation we know the least possible, i.~e.~only the two-point correlation function, equivalent to the second moment of the statistics. Now, the maximum entropy principle of statistics \citep[e.g.][]{Caticha} states that if we search for such a probability distribution, it must be Gaussian. Of course, a priori, we might even have no information about the two-point correlation. Nevertheless, the data itself yields such information, which we can extract during the inference procedure. 

For the problem of reconstructing a Gaussian signal field with unknown covariance, an optimal solution to the inference problem (\ref{invprob2}) can actually be found analytically or at least approximatively in calculating the posterior mean $\left<s\right>_{\mathcal{P}(s|d)}$ of the signal. A number of methods have been derived to do this, e.g. the critical filter and variants thereof \citep{EnsWeig, EnsFrom, ECF, SmP} or approaches using the method of Gibbs sampling \citep{Jens, WandeltLine1, WandeltLine2}. 

Unfortunately, if we consider the second (2.) and third (3.) constraints from above more closely, we must come to the conclusion that Gaussian signal fields are inappropriate for our problem since they are neither positive definite nor strongly fluctuative over orders of magnitude in strength. 

We consider instead that the logarithm of our signal field is Gaussian. If $s$ is a Gaussian field, $I = \mathrm{e}^{s}$ exhibits all the desired properties (1-3). It is known as a log-normal field. If we adapt the data model (\ref{invprob2})

\begin{equation}
d = RI + n = R I_{0} \mathrm{e}^{s} + n \label{datamodel}
\end{equation}
we are now faced with a considerably more complicated, non-linear problem. The factor $I_{0}$ can be set to account for the right units, w.l.o.g., we set it to one for the rest of this work. 

The likelihood $\mathcal{P}(d|s)$ and the signal prior $\mathcal{P}(s)$ take the following form 

\begin{align}
\mathcal{P}(d|s) &= \mathcal{G}(d-R\mathrm{e}^{s},N) \notag\\
		 &= \frac{1}{\mathrm{det}(2 \pi N)^{1/2}} \ \mathrm{e}^{-\frac{1}{2} \ (d-R\mathrm{e}^{s})^{\dagger} N^{-1} (d-R\mathrm{e}^{s})}, \\
\mathcal{P}(s)   &= \mathcal{G}(s,S) \notag\\
		 &= \frac{1}{\mathrm{det}(2 \pi S)^{1/2}} \ \mathrm{e}^{-\frac{1}{2} \ s^{\dagger} S^{-1} s}.
\end{align}
Then, the posterior of $s$ 

\begin{align}
\mathcal{P}(s|d) \propto \mathcal{G}(d-R\mathrm{e}^{s},N) \ \mathcal{G}(s,S) \label{posterior}
\end{align}
possibly becomes highly non-Gaussian due to the non-linearity introduced by (\ref{datamodel}).

Indeed, the resulting problem cannot be solved analytically. A possible approach would be to separate the quadratic and higher terms in (\ref{posterior}) 

\begin{align}
\mathcal{P}(s|d) \propto  \mathrm{e}^{-1/2 \ s^{\dagger}\left(S^{-1}+M\right)s \ + \ s^{\dagger}j \ + \ \sum\limits_{n=3}^{\infty} \Lambda^{n}_{x_{1} \cdots x_{n}} s_{x_{1}} \cdots x_{n}}  \label{FullIFT}
\end{align}
where $\Lambda^{n}$ is a rank -- $n$ tensor, and

\begin{align}
j &= R^{\dagger}N^{-1}d \\
M &= R^{\dagger}N^{-1} R.
\end{align}
The higher order terms could be handled either by invoking perturbative methods as known in statistical or quantum field theory \citep{Huang, PeskinSchroeder}, and already further developed for statistical inference \citep[e.g.][]{IFT}, or by using a Monte Carlo Gibbs sampling method \citep{MetroHast,GibbsSamp,HamiltonianSamp}. Since these methods are computationally very expensive for this log-normal ansatz and the high dimensionality of the problem, we do not follow them any further in this work. 

Instead, we seek an approximate solution in the signal field that maximizes the posterior:
\begin{equation}
\left<s\right>_{\mathcal{P}(s|d)} \approx \mathrm{argmax}_{s} \mathcal{P}(s|d). \label{posmean2map}
\end{equation}
This method is known as \textit{Maximum a posteriori} (MAP) in statistical inference\footnote{The maximum a posteriori approach can also be interpreted as an approximation to the posterior mean $\left<s\right>_{\mathcal{P}(s|d)}$, but is not guaranteed to yield a close result, especially not for highly non-Gaussian posterior shapes. Alternatively, it can be derived by minimizing an $\mathcal{L}_{\infty}$-norm error measure instead of the $\mathcal{L}_{2}$ minimization underlying the posterior mean approach.}. For the present problem it leads to a non-linear optimization problem of a gradient equation for the posterior. With this approach, it is further possible to calculate a consistent uncertainty estimate. In principle, the uncertainty of a signal reconstruction can be estimated by the width of the posterior. In this case, we use the inverse curvature of the posterior at its maximum to approximate the relative uncertainty $D$ (see App. \ref{appendix1} for details).

In this context, we still need to specify how to deal with the unknown correlation structure, i.e. the Gaussian signal covariance $S = \left<ss^{\dagger}\right>$. As mentioned earlier, the problem of reconstructing a Gaussian random field with unknown covariance has been solved already \citep{Jens, EnsWeig, EnsFrom, ECF, WandeltLine1}, and even the respective problem for a log-normal random field has been partly solved before \citep{SmP}. Unfortunately, none of these methods can be readily applied to the inference problem at hand, since they require the signal response to have a diagonal representation in signal space. This is not necessarily fullfilled for the Fourier-response (\ref{response}). We therefore develop a different approach, which nevertheless closely follows the previously mentioned works.

Crucially, as explained above, our prior knowledge signal statistics is homogeneous and isotropic. This implies that the unknown signal covariance becomes diagonal in its conjugate Fourier space and can be expressed by its power spectrum $P_{s}(|k|)$ \citep[see the Wiener-Kinchin theorem in][]{Bracewell}

\begin{align}
S(k,k') = \left<s(k)s(k')^{\dagger}\right> = (2\pi)^{n_{s}} \delta(k - k') P_{s}(|k|) \label{PS}
\end{align}
where $P_{s}(|k|)$ is just the Fourier transformation of the homogeneous and isotropic autocorrelation function $C(r) = S(|x-y|)$ 

\begin{equation}
 P_{s}(|k|) = \int dr \ C(r) \ \exp(ikr).
\end{equation}
We note that due to the assumption of isotropy, the power spectrum only depends on the length $|k|$ of the Fourier vector $k$. It is therefore sensitive to scales but not to full modes in Fourier space. Where the distinction is needed, we will make it explicit using the notation $|k|$.

We now parameterize the unknown covariance $S$ as a decomposition into spectral parameters $p_i$ and positive, disjoint projection operators $S^{(i)}$ onto a number of spectral bands such that the bands fill the complete Fourier domain 
\begin{equation}
S = \sum_{i} p_i S^{(i)}. 
\end{equation}     
These parameters can be introduced into the inference problem as a second set of fields to infer.

\begin{figure}
\includegraphics{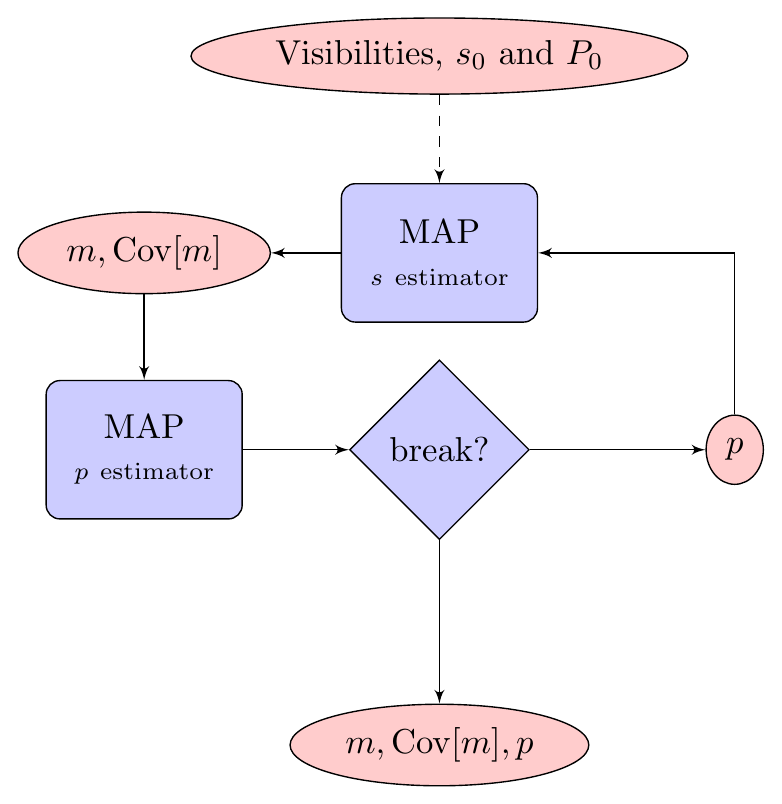}
\caption{Flow chart, illustrating the basic workflow of the \textsc{resolve} algorithm}. 
\label{fig:flow chart}
\end{figure}

We therefore add a second MAP algorithm to the signal MAP, solving for these unknown parameters $p_i$. We then iterate between both solvers until convergence is achieved. The algorithm produces a signal estimate $m$, an approximation to the reconstruction uncertainty $D$, and a power spectrum estimate parameter set $p_i$. The equations to be solved iteratively are

\begin{align}
&S^{-1}_{p} m + \mathrm{e}^m \cdot M \mathrm{e}^m - j \cdot \mathrm{e}^m = 0 \label{eq:one}\\ 
\notag\\
&\left(D\right)_{xy} = S^{-1}_{p \ xy} +  \mathrm{e}^{m_x} M_{xy} \mathrm{e}^{m_y} \notag\\
& \ \ \ \ \ \ \ \ \ + \mathrm{e}^{m_y} \int dz \ M(x,z) \ \mathrm{e}^{m(z)} \notag\\
& \ \ \ \ \ \ \ \ \ - j_x \cdot \mathrm{e}^{s_x} \ \delta_{xy} \label{eq:two} \\
\notag\\
&p_i = \frac{q_i + \frac{1}{2} \mathrm{tr} \left[(mm^{\dagger} + D)S^{(i)}\right]}{\alpha_i - 1 + \frac{\varrho_i}{2} + (Tp)_i} \label{eq:three}.
\end{align}
A detailed derivation can be found in App.~\ref{appendix1}. The two quantities $j$ and $M$ are defined as above, $q$ and $\alpha$ are parameters of a power spectrum parameter prior, $\varrho$ is a measure for the number of degrees of freedom of each Fourier band, and $T$ is an operator, which enforces a smooth solution of the power spectrum $p_i$. A thorough explanation of all these terms can be found in App.~\ref{appendix1}. Eq. (\ref{eq:one}) is the fix point equation that needs to be solved numerically to find a Maximum a Posteriori signal estimate $m$ for the current iteration. The second equation (\ref{eq:two}) results from calculating the second derivative of the posterior for the signal estimate $m$, its inverse serves as an approximation to the signal uncertainty $D$ at each iteration step. The last equation (\ref{eq:three}) represents an estimate for the signal power spectrum (and therefore its autocorrelation function), using the signal uncertainty $D$ to correct for missing signal power in the current estimate $m$. The iteration is stopped after a suitable convergence criterion is met (see App.~\ref{appendix2}). The whole algorithm is visualized in a flow chart in Fig.~\ref{fig:flow chart}.

It should be noted that solving these equations can be relatively time-consuming compared to e.g. MS-CLEAN, depending on the complexity of the problem at hand, since it involves a non-linear optimization scheme (\ref{eq:one}) and the numerical inversion and random probing of an implicitly defined matrix (\ref{eq:two})\footnote{The overall computational costs go roughly with $N_{\mathrm{global}} N_{\mathrm{pr}} O(\sqrt{n_s} n_d)$ in the limit of a large number of visibility measurements $n_d$. The $n_s$ are the number of pixels in image space, $N_{\mathrm{pr}}$ is the number of used random probing vectors to estimate matrix traces, and $N_{\mathrm{global}}$ is the global number of iterations \textsc{resolve} needs to converge (see App.~\ref{appendix2})} (for details, see App.~\ref{appendix2}).

We call the combined algorithm \textsc{resolve} (\textbf{R}adio \textbf{E}xtended \textbf{SO}urces \textbf{L}ognormal decon\textbf{V}olution \textbf{E}stimator).

Since the most severe problem in radio imaging is effectively how to extrapolate into unmeasured regions in $uv$-space to deconvolve the dirty beam from the dirty image (see Eq. \ref{invprob}), an explanation is in order of how \textsc{resolve} achieves this deconvolution. 

In the fix-point equation (\ref{eq:one}) to calculate the signal estimate $m$, the multiplicative term $\mathrm{e}^m$ acts as an effective convolution beam in Fourier space. Regions, where it has a significant value require only little modifications through the iterations in order to explain features in the data entering the equation via $j$. In contrast, regions where $m$ is very negative require a drastic modification to capture data features. Therefore, sidelobe structures of the dirty image are more comfortably accounted for by restructuring the existing stronger emission regions than by enforcing weaker sidelobes structures in the final signal estimate. By concentrating the resolved structures into the strong emission regions, the lognormal model extrapolates information in $uv$-space. The multiplicative $\mathrm{e}^m$ term acts as a convolution kernel in Fourier space, enforcing some amount of smoothness in the visibility structures. This smoothness is exploited by \textsc{resolve} for extrapolating the measured visibilities into the regions of $uv$-space without direct measurements. In this way, \textsc{resolve} is also capable of achieving some degree of superresolution by extrapolating beyond the largest visibilities.         

\section{Test Simulations}

In what follows, we present a range of tests of \textsc{resolve} using simulated data. We have implemented the algorithm\footnote{To get access to the code prior to its envisaged public release, please contact henrikju@mpa-garching.mpg.de or ensslin@mpa-garching.mpg.de.} in \textsc{Python} using the versatile signal inference library \textsc{NIFTy} \citep{nifty}. For all details of the implementation, we refer the reader to Sec.~\ref{SecTwo} and App.~\ref{appendix2}. We also show comparisons to CLEAN and MEM to benchmark the performance and fidelity of our algorithm.

\begin{figure}[h!]
\centering
  \subfigure[$uv$-coverage in units of \# of wavelengths.]{
    \includegraphics[width=0.45\textwidth]{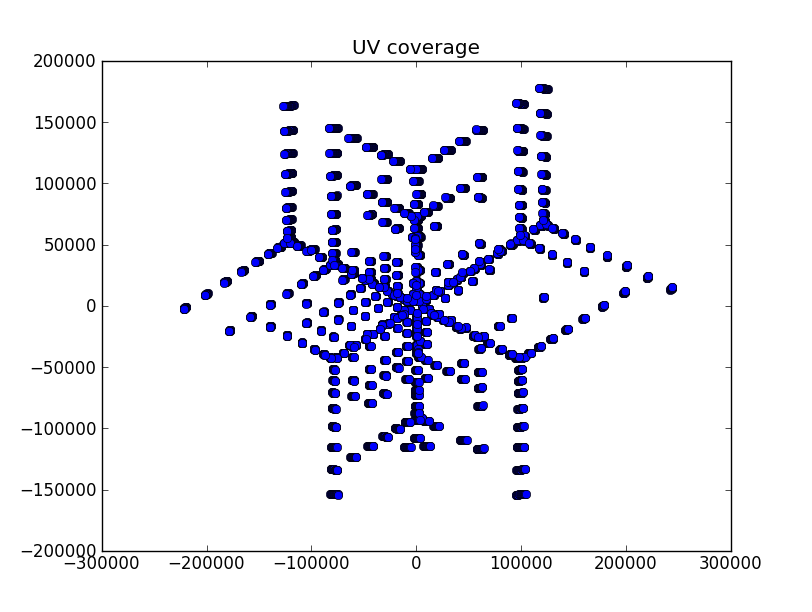}}
  \subfigure[Point spread function.]{
    \includegraphics[width=0.45\textwidth]{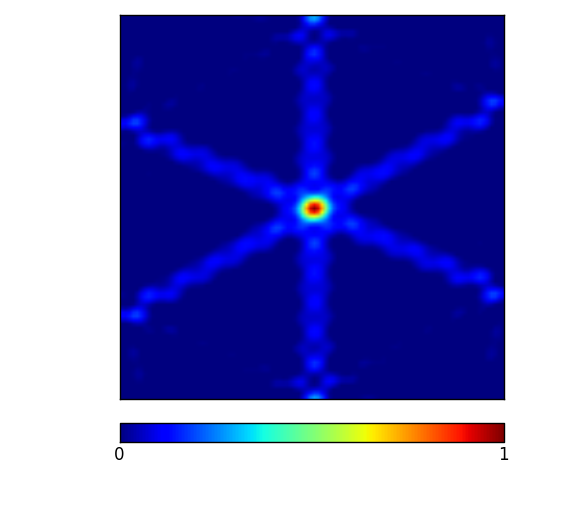}}
  \caption{$uv$-coverage and point spread function for the simulated 20 minutes snapshot observation in VLA-a configuration. The image of the point spread function is $100^2$ pixels large, the pixel size corresponds to roughly $0.2$ arcsec.} 
\label{realuvcov}
\end{figure}

For all tests, we constructed simulated observations with the tool \textsc{makems}\footnote{See http://www.lofar.org/wiki/lib/exe/fetch.php \\ ?media=software:makems.pdf.} using a realistic $uv$-coverage from a VLA observation in its A-Configuration. The VLA samples the $uv$-plane non-uniformly at irregular intervals, and the response includes thereby a convolutional gridding and degridding operator using a Kaiser-Bessel kernel (for details see Eq.~\ref{gridresponse} and App.~\ref{appendix2}). We simulated observations at a single frequency, approximatively 20 minutes snapshot observation with a total of 42\,120 visibility measurements at a central frequency of 1 GHz (see Fig. \ref{realuvcov}). This setting leads to an especially sparse sampling of the $uv$-plane. For ease of code development and testing, we have not used longer observations. On the other hand, if we can solve the more demanding cases of sparse $uv$-coverage, we certainly can handle better suited data. 

Through all simulations, we varied thermal visibility noise levels and input signals. 

For the next two sections (\ref{mainres} and \ref{SubSec: CLEANaMEM}), the signals were drawn from a log-normal distribution, exactly meeting our prior assumptions. In Sec.~\ref{realmock}, we go beyond that and illustrate the validity of our statistical model by using a signal derived from a CLEAN image of a real source.

The complex, Gaussian input noise in $uv$-space is defined by a user-defined variance, equal for all visibilities. The code does not require equal noise variances and can in principle handle varying variances as well. Always in the following, \textit{low noise} refers to $\sigma_{\mathrm{ln}}^2=10^{-3} \mathrm{Jy}^2$, whereas \textit{high noise} denotes $\sigma_{\mathrm{hn}}^2=10^{5} \mathrm{Jy}^2$ \footnote{The unit Jy was used here for convenience. Effectively, it stands for whatever units the simulated signal is interpreted to be given in.}. These numbers are of course somewhat arbitrary, only chosen for demonstrational reasons. They are not intended to necessarily reflect realistic visibility noise values in every possible aspect, but to serve as examples for particularly low or high noise cases.

To give a quantitative account of the accuracy of the reconstructions, we use a relative $\mathcal{L}_{2}$ - norm measure of the difference of signal to map:

\begin{equation}
\delta = \sqrt{\left(\frac{\sum\left(\mathrm{e}^s - \mathrm{e}^m\right)^2}{\sum \left(\mathrm{e}^{s}\right)^2}\right)}
\end{equation}
where the sums are taken over all pixels of the reconstruction. This choice is motivated by the fact that the inference approach underlying \textsc{resolve} approximates a reconstruction that is optimal in the sense of minimizing this error measure (see Sec.~\ref{SecTwo} and Eq.~\ref{posmean2map} therein).

In Secs. \ref{mainres} - \ref{realmock}, we focus exclusively on the reconstruction of the signal, i.e. the sky brightness distribution. The reconstruction of the power spectrum is discussed separately in Sec. \ref{SecPow}. 

\subsection{Main Test Results} \label{mainres}

Here, we describe the main test results for the reconstruction of a simulated signal using \textsc{resolve}.

\begin{figure*}[p]
   \centering
	\subfigure[Signal.]{
		\includegraphics[width=0.33\textwidth]{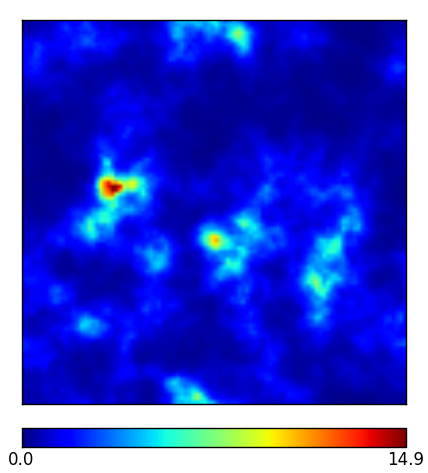}}
	\subfigure[Dirty Image.]{
		\includegraphics[width=0.33\textwidth]{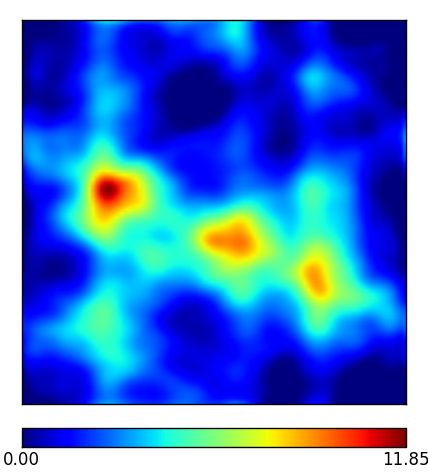}}\\
        \subfigure[\textsc{resolve} reconstruction with low noise.]{
		\includegraphics[width=0.33\textwidth]{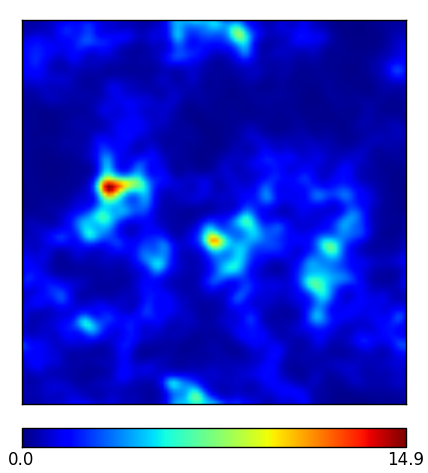}}
	\subfigure[Absolute error $\left|\mathrm{e}^s-\mathrm{e}^m\right|$.]{
		\includegraphics[width=0.33\textwidth]{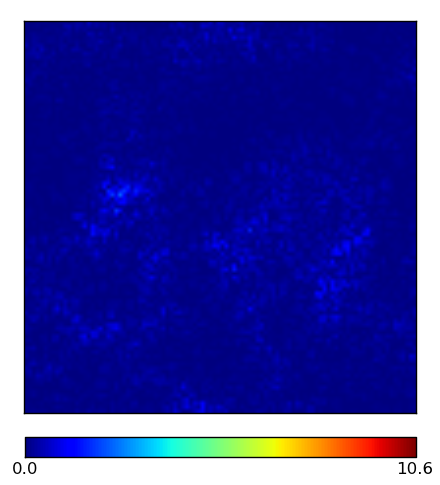}} \\
	\subfigure[\textsc{resolve} reconstruction with high noise.]{
		\includegraphics[width=0.33\textwidth]{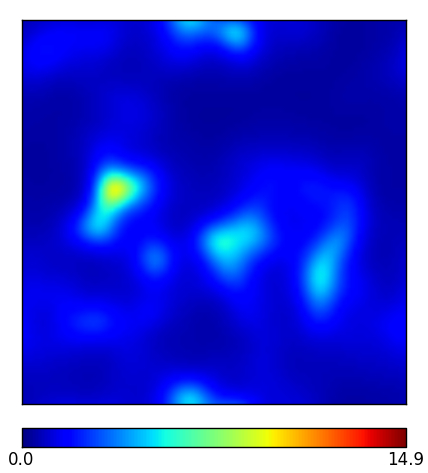}}
	\subfigure[Absolute error $\left|\mathrm{e}^s-\mathrm{e}^m\right|$.]{
		\includegraphics[width=0.33\textwidth]{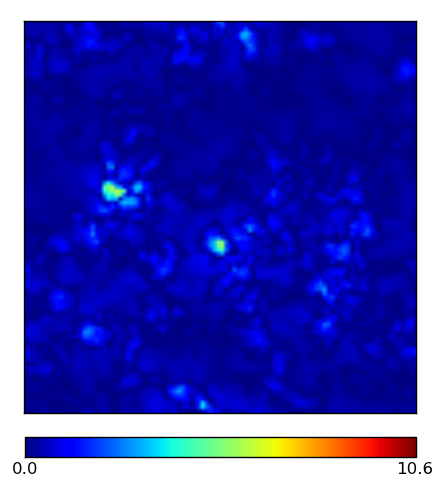}}\\
	\caption{Reconstruction of a log-normal signal field, observed with a sparse $uv$-coverage from a VLA-A-configuration \label{realRecon1} and different noise levels. The images are $100^2$ pixels large, the pixel size corresponds to roughly $0.2$ arcsec. The brightness units are in Jy/px. The ridge-like structures  in the difference maps simply stem from taking the absolute value and mark zero-crossings between positive and negative errors. \textit{First row left}: Signal field. \textit{First row right}: Dirty map.\textit{Second row left} \textsc{resolve} reconstruction with low noise. \textit{Second row right}: Absolute per-pixel difference between the signal and the \textsc{resolve} reconstruction with low noise. \textit{Third row left}: \textsc{resolve} reconstruction with high noise. \textit{Third row right}: Absolute per-pixel difference between the signal and the \textsc{resolve} reconstruction with high noise.}
\label{lownoiseRecon1}  
\end{figure*}

In Fig. \ref{lownoiseRecon1}, an artificial log-normal signal is shown alongside with the results from \textsc{resolve} for observations with low and high noise. The error measures are $\delta_{\textrm{ln}} = 0.12$ and $\delta_{\textrm{hn}}= 0.3$ for the low and high noise case respectively.

We see that we can recover all the structures of the original surface brightness, down to even very small features in the low noise case and at least all main features in the high noise case. All strong effects of the point spread function have been successfully removed, thus showing that \textsc{resolve} is effective in deconvolving the dirty image.

In fact, the reconstruction is expected to be smoothed out on the smaller scales and lose overall power due to noise in the observation. This is simply because all information in the power spectrum gets lost for powers comparable to the noise variance (see Sec. \ref{SecPow} for details). Effectively, \textsc{resolve} performs an automatic visibilty weighting by comparing noise and signal power for all Fourier modes, much in the way robust weighting originally was conceived \citep{BriggsThesis, BriggsPaper}. For a detailed discussion of this topic see App. \ref{appendix3}.

\subsection{Comparison to standard imaging methods}\label{SubSec: CLEANaMEM}

In this section, we give a short introduction to common imaging algorithms in radio interferometry and show comparisons to \textsc{resolve}. We focus on two of them, MS-CLEAN and MEM, which are probably the most widespread methods to date. 

In addition, we should mention recent developments in the application of Compressed Sensing (CS) \citep{CS1,CS2} to radio imaging, most notably the development of the imaging algorithm SARA \citep{SARA}. Another approach was taken recently to apply Gibbs sampling methods to imaging in radio interferometry \citep{SutterWandeltRadio}, also within the framework of Bayesian inference, but restricted to pure Gaussian priors. A direct comparison of \textsc{resolve} to either SARA or Gibbs Sampling methods is out of the scope of this work, but we will discuss possible ways to include the CS approach into our Bayesian framework in the conclusions (see Sec. \ref{Sec: Conclusions}).

Standard imaging methods almost always use an additional visibility weighting that is manually set by the observer. We do not include such weights explicitly in \textsc{resolve} (i.e.~we set them to one). However, as already mentioned, an algorithm like \textsc{resolve} \textit{automatically} chooses optimal weights according to the ratio of reconstructed source power to noise power. A detailed derivation can be found in App.~\ref{appendix3}. For CLEAN, we compare to different weighting schemes in order to be as unbiased as possible.

For both CLEAN and MEM we used the implementation in the radio astronomical software package CASA \citep{CASA}.

\subsubsection{Comparison to CLEAN}

CLEAN was first presented by \citep{1974A&AS...15..417H} and is surely the most widely used deconvolution algorithm in radio astronomy. It works around the major assumption that the image is comprised of point sources. In its simplest variant, it iteratively finds the highest peak in the dirty map, subtracts a psf-convolved fraction of a delta function fitted to the peak, and saves the delta components in a separate image. After some noise threshold is reached, the algorithm stops and reconvolves the components with a so called clean beam, usually the main lobe of the point spread function or a broader version of it to downgrade resolution. 

Over time, many variants of CLEAN have been developed, most notably Clark CLEAN \citep{ClarkClean}, Cotton-Schwab CLEAN \citep{CSClean}, multifrequency CLEAN \citep{MFClean} and multiscale CLEAN (MS-CLEAN) \citep{MSClean}. The latter was constructed to better reflect extended emission by subtracting Gaussians of various shapes instead of pure point sources. We will thus compare the results of \textsc{resolve} to MS-CLEAN.

It has been proved in the framework of compressed sensing that CLEAN is in fact a variant of a matching pursuit algorithm \citep{WipeClean}. This class of algorithms can be shown to be optimal for signals that are sparse in some basis (like a point source signal on the sky). It can be understood as minimizing the $\mathcal{L}_1$-norm of the signal field and can therefore be cast into the assumption of a Laplacian prior distribution, which would allow in principle a representation of CLEAN in a Bayesian inference framework \citep[see][and references therein]{WiauxScaifePaper}. 

\begin{figure*}[p]
   \centering
	\subfigure[\textsc{resolve} reconstruction.]{
		\includegraphics[width=0.44\textwidth]{m_n1e-3_crop.png}}
	\subfigure[Absolute error $\left|\mathrm{e}^s-\mathrm{e}^m\right|$.]{
		\includegraphics[width=0.44\textwidth]{m_n1e-3_absdiff_crop.png}}\\
	\subfigure[CLEAN map with natural weighting.]{
		\includegraphics[width=0.44\textwidth]{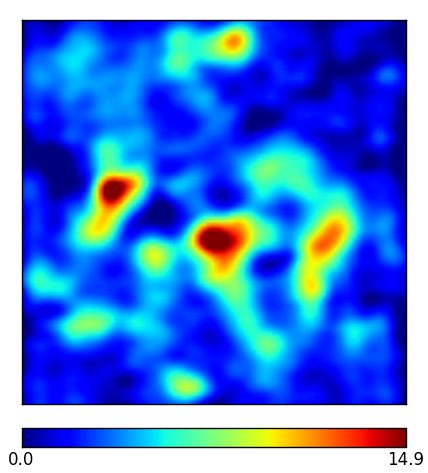}}
	\subfigure[Absolute error $\left|\mathrm{e}^s-m_{\mathrm{natural}}\right|$.]{
		\includegraphics[width=0.44\textwidth]{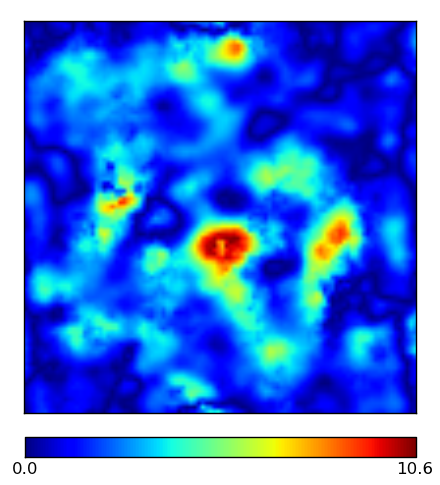}}\\
	\vspace{0cm}
	\caption{Comparison of \textsc{resolve} with MS-CLEAN for the simulated low noise observation of Sec.~\ref{mainres}. The images are $100^2$ pixels large, the pixel size corresponds to roughly $0.2$ arcsec. The brightness units are in Jy/px. The ridge-like structures simply stem from taking the absolute value and mark zero-crossings between positive and negative errors. \textit{First row left}: \textsc{resolve} reconstruction. \textit{First row right}: Absolute per-pixel difference between the signal and the \textsc{resolve} reconstruction with. \textit{Second row left}: MS-CLEAN reconstruction with natural weighting using the radio astronomical software package CASA. \textit{Second row right}:  Absolute per-pixel difference between the signal and the MS-CLEAN reconstruction with natural weighting.}
	\vspace{4cm}
	\label{cleancomparison1}
\end{figure*}

\begin{figure*}[p]
   \centering
	\subfigure[CLEAN map with uniform weighting.]{
		\includegraphics[width=0.44\textwidth]{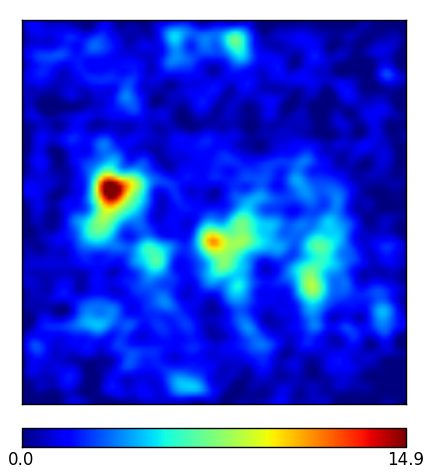}}
	\subfigure[Absolute error $\left|\mathrm{e}^s-m_{\mathrm{uniform}}\right|$.]{
		\includegraphics[width=0.44\textwidth]{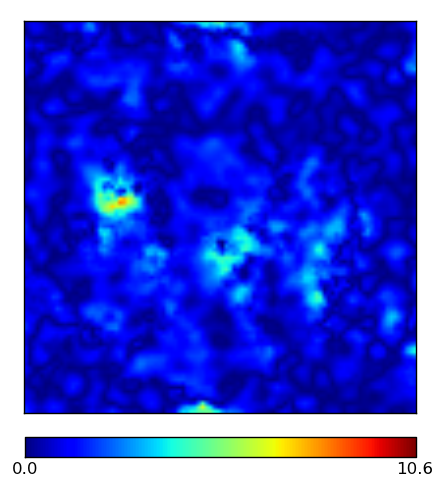}}\\
	\subfigure[CLEAN map with robust weighting.]{
		\includegraphics[width=0.44\textwidth]{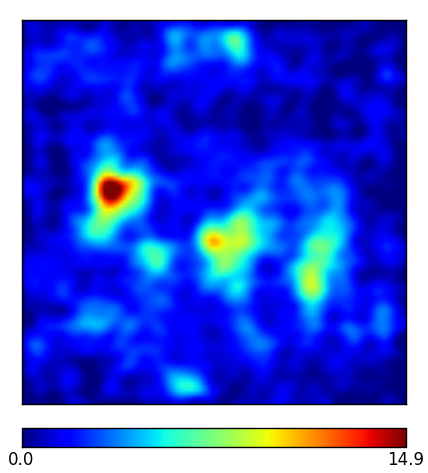}}
	\subfigure[Absolute error $\left|\mathrm{e}^s-m_{\mathrm{robust}}\right|$.]{
		\includegraphics[width=0.44\textwidth]{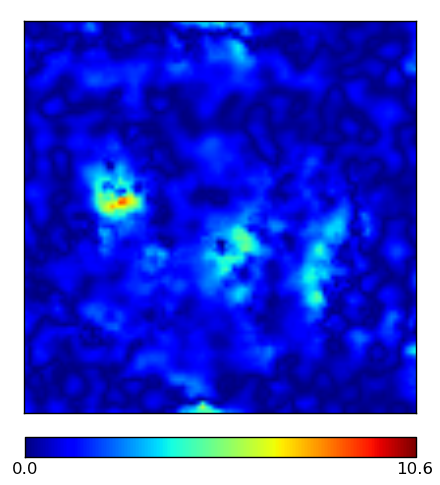}}\\
	\vspace{0cm}
	\caption{Comparison of \textsc{resolve} with MS-CLEAN for the simulated low noise observation of Sec.~\ref{mainres}. The images are $100^2$ pixels large, the pixel size corresponds to roughly $0.2$ arcsec. The brightness units are in Jy/px. The ridge-like structures simply stem from taking the absolute value and mark zero-crossings between positive and negative errors.  \textit{First row left}: MS-CLEAN reconstruction with uniform weighting using the radio astronomical software package CASA.\textit{First row right}:  Absolute per-pixel difference between the signal and the MS-CLEAN reconstruction with uniform weighting. \textit{Second row left}: MS-CLEAN reconstruction with robust ($r=0$) weighting using the radio astronomical software package CASA. \textit{Second row right}: Absolute per-pixel difference between the signal and the MS-CLEAN reconstruction with uniform weighting.}
	\vspace{4cm}
	\label{cleancomparison2}
  \end{figure*}

In Figs.~\ref{cleancomparison1} and \ref{cleancomparison2}, a comparison is shown between the results of \textsc{resolve} and MS-CLEAN as implemented in the radio astronomical software package CASA. For this test, the same simulated low noise data were used as in Sec.~\ref{mainres}. We compare to three different CLEAN reconstructions with natural, uniform and robust weighting (robust parameter $r=0$, which gives an intermediate result between the other two schemes). We used a very small noise threshold and a standard gain factor of $0.1$. In total, we choose to run the algorithm interactively for around 1000 iterations. We used around ten different scales for the multi-scale settings, ranging from a few pixels to enough to roughly match the scales found in the signal. Together with the reconstructions, we show maps of the squared error $(\mathrm{e}^s-m)^2$ for each of them. The $\mathcal{L}_2$ - error measures are shown in Table \ref{tab:l2clean}. 

Both quantitative analysis and visual comparison show that \textsc{resolve} clearly outperforms MS-CLEAN in this case. Its result is closer to the signal in the $\mathcal{L}_2$
error measure sense and it is clearly superior in reconstructing the detailed extended structure of the surface brightness signal. Especially the
very weak emission around all the brighter sources is much better resolved and denoised than in the MS-CLEAN images. The reconstruction with
natural weighting is overestimating the flux scales considerably, while uniform and robust weighting roughly find the same correct solution as
\textsc{resolve}. However, it should be noted that, at least for natural weighting, this is a somewhat biased comparison, since the natural weighting scheme is by construction enhancing point-source sensitivity while preserving larger side-lobe structures \citep{BriggsThesis} and thus not the optimal choice for resolving extended emission.

\begin{table}
\centering
\begin{tabular}{l|c}
Algorithm & $\delta$ \\
\hline
\textsc{resolve} & 0.12\\
MS-CLEAN, natural & 1.46\\
MS-CLEAN, uniform & 0.67\\
MS-CLEAN, robust & 0.69\\
MEM & 1.07
\end{tabular}
\caption{$\mathcal{L}_2$ error measures for \textsc{resolve}, MS-CLEAN and MEM for the low-noise simulation and the reconstruction shown in Figs.~\ref{cleancomparison1}, \ref{cleancomparison2} and \ref{memcomparison}.}
\label{tab:l2clean}
\end{table}

\subsubsection{Comparison to the Maximum Entropy Method (MEM)} 

The maximum entropy method (MEM) is an imaging algorithm introduced into radio astronomy by \citep{MEM}. It actually goes back to earlier developments in statistical inference, connected to the broad field of entropic priors \citep{GullMEM,SkillingMEM}. It should not been confused with the maximum entropy principle of statistics \citep[][see also Sec. \ref{Sec:MainAlgo}]{Caticha,EnsWeig} that describes how to update probability distributions when new information has to be included. 

MEM aims to maximize a quantity called image entropy $S_{\mathrm{im}}$, which is defined for strictly positive signal images $s$ as

\begin{equation}
S_{\mathrm{im}} = - \int dx \ s(x) \log \left(s(x)/m(x)\right) \label{ImEnt}
\end{equation}
where $m(x)$ is a model image of the observed signal, thus allowing to introduce some kind of prior information into the problem. The data enter this formalism as a constraint for the maximization problem. Usually, one adds a term to (\ref{ImEnt}) that measures the closeness of the entropic signal reconstruction to the data in the form of a $\chi^{2}(d,Rs)$ distribution, which is nothing else but the log-likelihood of (\ref{likeli2}):

\begin{align}
\frac{1}{2} \chi^{2} (d,Rs) &= \frac{1}{2} (d-Rs)^{\dagger} N^{-1}(d-Rs) \notag\\
		&= -\log(P(d|s)) + \mathrm{const}. \label{Chi2Ent}
\end{align}

With (\ref{ImEnt}) and (\ref{Chi2Ent}), MEM achieves a solution by extremizing

\begin{align}
J(d,s) = -\log{P(d|s)} - \mu S_{\mathrm{im}}
\end{align}
for $s$. The multiplier $\mu$ is usually adjusted during the extremization to meet numerical constraints \citep[see][for details]{MEM}. 

We now repeat a short section from \citep{EnsWeig}, analyzing the assumptions of this approach from the viewpoint of Bayesian signal inference.

As we have identified (\ref{likeli2}) as the log-likelihood, it is also possible to re-identify the prior distribution. If we interpret $J(d,s)$ as a Hamiltonian $H(d,s)$, than the entropy term can be understood as a log-prior

\begin{align}
\mu S_{\mathrm{Im}}(s) = \log \mathcal{P}(s).
\end{align}
With this, we can read off the underlying prior distribution implicitly assumed in MEM
\begin{align}
\mathcal{P}(s) &= \exp{\left[- \mu \int \mathrm{d}x s(x) \log\left(\frac{s(x)}{m(x)}\right)\right]} \notag\\
  	       &= \prod_{x} \left(\frac{s(x)}{m(x)}\right)^{-\mu s(x)}.
\end{align}
This prior is very specific. It extremely suppresses strong pixel values and thereby favors to smooth out emission over all pixels in the image while sharp peaks are heavily down-weighted. It implicitly assumes no correlation between pixels, and a more than exponentially falling brightness distribution. In the case of the model $m(x)$ being a close approximation to real signal, the prior becomes effectively flat and MEM turns basically into a Maximum Likelihood fit. 

\begin{figure*}[ph!]
   \centering
	\subfigure[\textsc{resolve} reconstruction.]{
		\includegraphics[width=0.44\textwidth]{m_n1e-3_crop.png}}
	\subfigure[Absolute error $\left|\mathrm{e}^s-\mathrm{e}^m\right|$.]{
		\includegraphics[width=0.44\textwidth]{m_n1e-3_absdiff_crop.png}}\\
	\subfigure[MEM map.]{
		\includegraphics[width=0.44\textwidth]{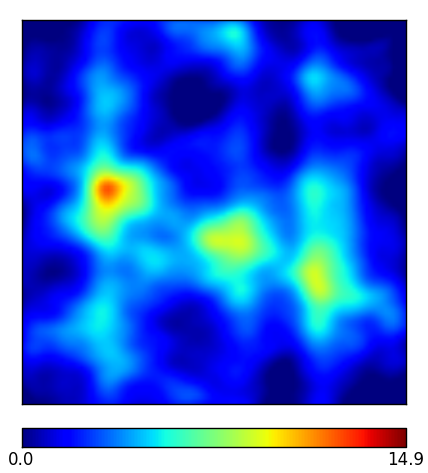}}
	\subfigure[Absolute error $\left|\mathrm{e}^s-m_{\mathrm{MEM}}\right|$.]{
		\includegraphics[width=0.44\textwidth]{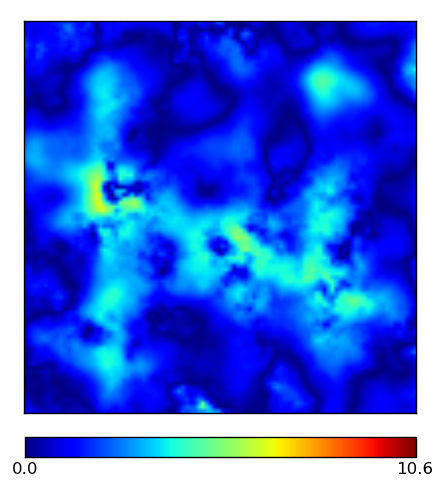}}\\
	\caption{Comparison of \textsc{resolve} with MEM for the simulated low noise observation of Sec. \ref{mainres}. The images are $100^2$ pixels large, the pixel size corresponds to roughly $0.2$ arcsec. The brightness units are in Jy/px. The ridge-like structures simply stem from taking the absolute value and mark zero-crossings between positive and negative errors.  \textit{First row left}: \textsc{resolve} reconstruction. \textit{First row right}: Absolute per-pixel difference between the signal and the \textsc{resolve} reconstruction. \textit{Second row left}: MEM reconstruction using the radio astronomical software package CASA. \textit{Second row right}: Absolute per-pixel difference between the signal and the MEM reconstruction.}
	\vspace{4cm}
\label{memcomparison}  
\end{figure*}

In Fig.~\ref{memcomparison}, a comparison is shown between the results of \textsc{resolve} and MEM as implemented in the radio astronomical software package CASA. Again, the same simulated low noise data were used as in Sec. \ref{mainres}. As a model image, we used an MS-CLEAN reconstruction with uniform weighting. We again show maps of the squared error $(\mathrm{e}^{s}-m)^2$ for the reconstruction with \textsc{resolve} and MEM respectively. The $\mathcal{L}_2$ error measures are shown in Table \ref{tab:l2clean}.


It can be clearly seen that \textsc{resolve} also outperforms MEM. There is much more false structure in the MEM reconstruction, as reflected by the $\ell_2$ - norm analysis. Partly, this might be due to the specific MEM prior that enforces to smooth out the signal over all pixels, partly, it seems to be due to badly deconvoled remnants of the point spread function. However, we note that the MEM implementation in CASA still is considered to be somewhat experimental, and that a more stable code or a longer time of parameter adjustment and fine-tuning might improve these results.

\subsection{Comparison with a real signal}\label{realmock}

\begin{figure*}[p]
   \centering
	\subfigure[\textsc{resolve} reconstruction.]{
		\includegraphics[width=0.44\textwidth]{m_n1e-3_crop.png}}
	\subfigure[Absolute error $\left|\mathrm{e}^s-\mathrm{e}^m\right|$.]{
		\includegraphics[width=0.44\textwidth]{m_n1e-3_absdiff_crop.png}}\\
	\subfigure[Relative uncertainty map.]{
		\includegraphics[width=0.44\textwidth]{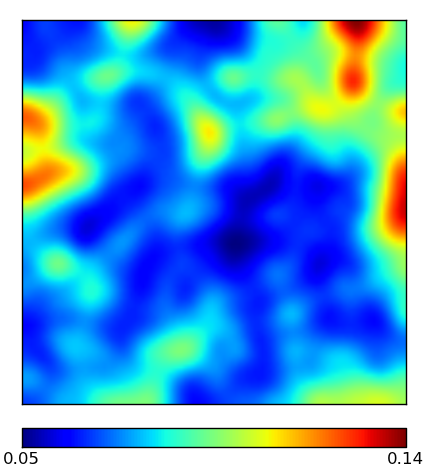}}
	\subfigure[Relative difference map $\left|\mathrm{e}^s-\mathrm{e}^m\right| / \mathrm{e}^s$]{
		\includegraphics[width=0.44\textwidth]{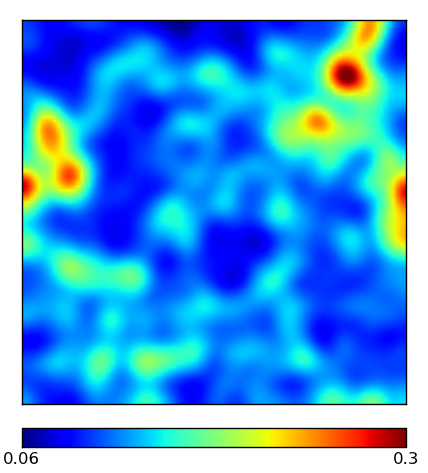}}\\
	\caption{\textit{First row left}: \textsc{resolve} reconstruction for the low noise reconstruction of Sec. \ref{mainres}. \textit{First row right}: Absolute per-pixel difference between the signal and the \textsc{resolve} reconstruction. The ridge-like structures simply stem from taking the absolute value and mark zero-crossings between positive and negative errors. \textit{Second row left}: Relative Uncertainty map derived from the \textsc{resolve} reconstruction. \textit{Second row right}: Relative difference map between signal and \textsc{resolve} reconstruction. }
	\vspace{4cm}
\label{fig:uncertainty} 
\end{figure*}

So far we have only shown reconstructions of signals that were drawn from log-normal statistics, using the exact assumptions hat we use to specify the prior distribution. To some degree, it is expected that \textsc{resolve} should be optimal for these simulated signals.

\begin{figure}[h!]
   \centering
	\subfigure[Signal.]{
		\includegraphics[width=0.42\textwidth]{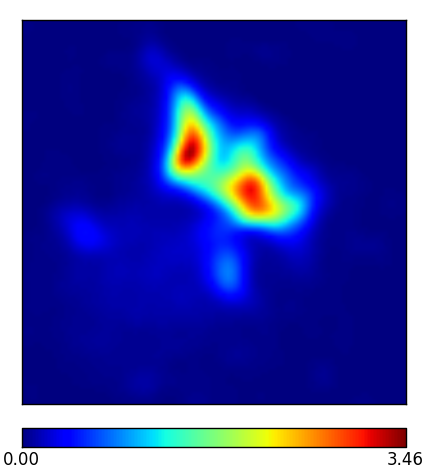}}\\
	\subfigure[\textsc{resolve} reconstruction.]{
		\includegraphics[width=0.42\textwidth]{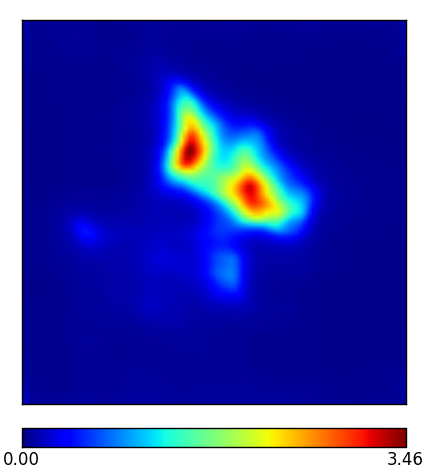}}\\
	\caption{Reconstruction of a signal field that was obtained from a CLEAN image of the real extended emission of Galaxy cluster Abell 2256. For the simulation, the same setup with low noise was used as in Sec. \ref{mainres}.}\label{realmockfig}
\end{figure}

To further demonstrate the validity of our assumptions, we have conducted a test where we did not use a signal drawn from log-normal statistics. Instead, we took an MS-CLEAN image, obtained from real data of the galaxy cluster Abell 2256 \citep{Tracy}, and reused this as a signal for the simulated observation using the same VLA configuration as before. The original data were taken with the VLA at $1.369$ GHz in D-configuration. The surface brightness values are not in the original range but chosen arbitrarily in our simulation, effectively given in Jy/px. The signal (i.e. the adapted CLEAN image of Abell 2256) and the reconstruction from \textsc{resolve} are shown in Fig.~\ref{realmockfig}. 

Although this time we have at no point introduced log-normal statistics into the simulation process, the prior assumption still seems to be valid and leads to results comparable in exactness to the tests using explicit log-normal signals.

%

\subsection{Signal Uncertainty}

As already stated in Sec. \ref{Sec:MainAlgo}, \textsc{resolve} provides also an estimate of the uncertainty of the signal reconstruction. The algorithm uses the inverse second derivative $D$ of the posterior, evaluated at the specific signal estimate $m$, to approximate the posterior covariance. In App.~\ref{UncertaintyTheory}, it is shown that a full signal estimate taking into account approximative uncertainty leads to

\begin{equation}
I \approx \mathrm{e}^{m_x} \pm \sqrt{\mathrm{e}^{2m_x} \left[\mathrm{e}^{D_{xx}} -1\right]}. \label{fullestimate2}
\end{equation}

In Fig. \ref{fig:uncertainty}, we present an example of the approximated relative uncertainty 

\begin{equation}
\sqrt{\frac{\langle \left(\mathrm{e}^{s_x}\right)^2 \rangle_{\mathcal{G}(m,D)} -  \langle \mathrm{e}^{s_x} \rangle_{\mathcal{G}(m,D)}^2}{\langle \mathrm{e}^{s_x} \rangle_{\mathcal{G}(m,D)}^2}} = \sqrt{\left[\mathrm{e}^{D_{xx}} -1\right]}
\end{equation}
for the low noise reconstruction of Sec.~\ref{mainres}, together with the signal estimate, and absolute and relative difference map between signal and estimate. The subscripts indicate that our approach effectively involves to approximate the full posterior with a Gaussian $\mathcal{G}(m,D)$ centered on the signal estimate and with a covariance of $D$ (see App.~\ref{UncertaintyTheory}).

Fig.~\ref{fig:uncertainty} shows that the uncertainty follows the structure of the reconstruction. Where the signal is strong, the relative uncertainty is much lower than in regions that are mainly dominated by noise. A comparison between the estimated relative uncertainty and the real relative difference map shows the approximative nature of the theoretical estimate. While both maps agree nicely in structure, they do not fully match in terms of values. Overall, the theoretical uncertainty underestimates the real relative difference. However, it should be noted that the deviations between both maps are much stronger in the outer regions, where the signal is only weak. In the center of the map, where the source mainly is located, both agree relatively well.

If we further use (\ref{fullestimate2}) to calculate the absolute uncertainty for the low noise reconstruction of Sec.~\ref{mainres}, we find that roughly $40 \ \%$ of the original signal values lie within a $1-\sigma$ region, and roughly $70 \ \%$ within a $2-\sigma$ region. Although this result deviates from pure Gaussian expectations, this is a reasonable outcome. Since the posterior is in general non-Gaussian, the assumption of posterior Gaussianity needed to exactly define (\ref{fullestimate2}) can only result in an approximation.

Calculating the uncertainty to a very high precision is computationally expensive\footnote{The estimation of the uncertainty goes roughly with $N_{\mathrm{pr}}\left(O(\sqrt{n_s} n_d) + O(\sqrt{n_s} n_s \log(n_s))\right)$, where $N_{\mathrm{pr}}$ is the number of probes, $n_d$ the number of visibility measurements, and $n_s$ the number of pixels in image space (see App.~\ref{appendix2})} . It involves the probing of an implicitly defined matrix and a numerical algorithm to invert this matrix (see App.~\ref{appendix2}). In this case, we have stopped the stochastic probing of $D$ at some point for computational reasons and smoothed the outcome a bit to obtain Fig.~\ref{fig:uncertainty}. This might add to the deviations from pure Gaussian expectations on the absolute uncertainty, mentioned earlier. Nevertheless, since the theoretical matrix representation of $D$ must be smooth, this procedure should be acceptable as long as this example simply serves as a showcase to fundamentally demonstrate how to obtain an uncertainty estimate with \textsc{resolve}.

\subsection{Power Spectrum Reconstructions} \label{SecPow}

Until now, we have focused entirely on the reconstruction of signal maps. Now we discuss the reconstruction of the signal power spectrum that \textsc{resolve} achieves automatically in order to infer the best signal solution. The signal power spectrum is defined as the Fourier transformation of the autocorrelation function of the signal, assuming translationally and rotationally invariant statistics:

\begin{equation}
 P(|k|) = \int \mathrm{d}r \ C(r) \ \exp(ikr).
\end{equation}
(for more details, see Sec. \ref{Sec:MainAlgo}).

Qualitatively, it can be understood as decomposing the signal autocorrelation into its different contributions from various scales. High power on low Fourier modes means strong correlations on larger scales and high power on high Fourier modes means strong correlations on smaller scales.

\begin{figure}[tb]
   \centering
	\subfigure{
		\includegraphics[width=0.5\textwidth]{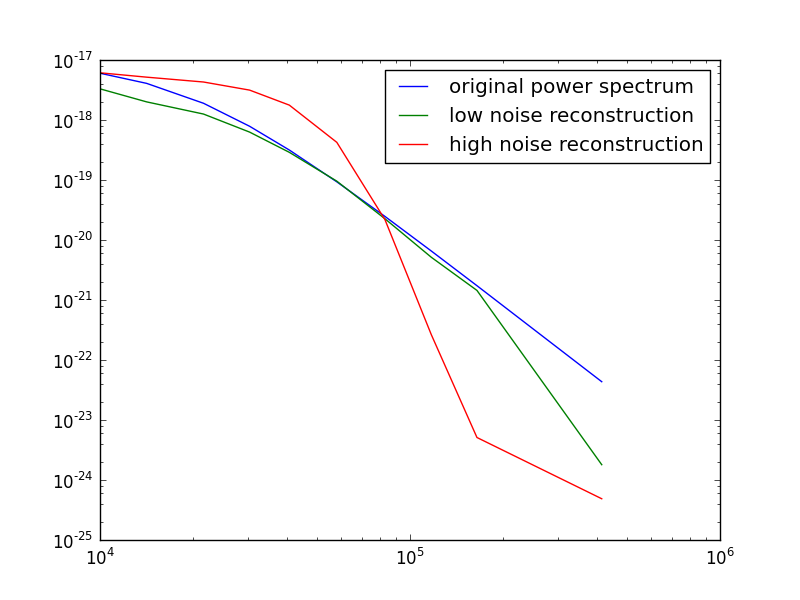}}
	\subfigure{
		\includegraphics[width=0.5\textwidth]{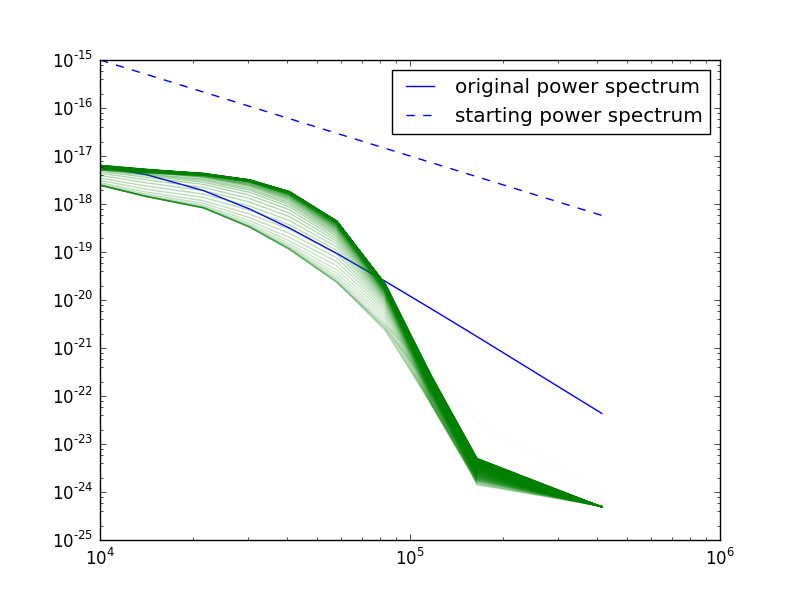}}
	\caption{\textit{First row}: Power spectrum reconstruction for the simulated low noise and high noise observations of Sec.~\ref{mainres}. \textit{Second Row}: Evolution of the high noise power spectrum reconstruction over 80 iterations. The iteration process is indicated from transparent to full green.}
\label{ps_total}
\end{figure}

In the first row of Fig.~\ref{ps_total}, we show the reconstruction of power spectra for the low and high noise reconstructions of Sec.~\ref{mainres}. The figure shows the original power spectrum, which defines the correlation structure of the signal field, and the final results of \textsc{resolve} after 6 iterations in the low, and 80 iterations in the high noise case. It can be seen that, with more noise, the reconstruction looses sensitivity for the smaller scales. This is reflected in the high noise map reconstruction in Fig.~\ref{lownoiseRecon1}, where the smallest scales are smoothed out by the algorithm.

The second row of Fig.~\ref{ps_total} serves as an example for the actual reconstruction process, where all of the 80 iterations for the high noise power spectrum are shown, together with the starting guess, which was a simple and generic power law $P_{\mathrm{sg}} \propto k^{-2}$. The power spectrum first dropped, to slowly rise up again. This is a consequence of a numerical procedure to ensure the convergence of the underlying non-linear optimization routines, where a constant diagonal is first added to the uncertainty estimate $D^{-1}$ used in the power spectrum reconstruction, and then suppressed again with converging iterations (see App.~\ref{appendix2}).

We emphasize that an accurate power spectrum reconstruction can be a scientific result on its own and should not only be regarded as a mere by-product. Since this is a rather unusual topic for observations of radio total intensity, it might be in place to explain a little further its meaning and to outline possible scientific merits. 

The most typical physical source of extended emission in radio astronomy is synchrotron radiation. By spelling the power spectrum of the total intensity from some astronomical synchrotron source we effectively measure its correlation structure. Since synchrotron intensity is in part determined by the magnetic field strength \citep{RL} in the source, we automatically gather valuable scientific information on the magnetic field statistics as well:

\begin{equation}
C_{I}(r) = \left<I(x)I(x+r)\right> \propto \left<B(x)^2B(x+r)^2\right>. \label{II}
\end{equation}
Detailed derivations of this and related statistical quantities, together with many discussions on its scientific use, mostly in the context of analyzing turbulent magnetic fields, can be found in a series of astrophysical papers \citep[e.g.][]{Spangler1, Spangler2, Eilek, Waelkens, Imprints, Probing, Lazarian}  

For future observations, it might be especially interesting to use these results from \textsc{resolve} to compare data of specific astrophysical synchrotron sources, e.g.~supernova remnants or radio halos of galaxies and clusters, to simulations thereof. In simulations, the inputs are under control, and (\ref{II}) can actually be calculated and compared with real data.

\section{Conclusions}\label{Sec: Conclusions}

In this paper we presented a new approach to signal inference and imaging in radio astronomy and especially radio interferometry. The inference algorithm \textsc{resolve} is targeted to be optimal for the imaging of extended and diffuse radio sources in total intensity. In simulations, \textsc{resolve} demonstrated to produce high fidelity reconstructions of such extended signals, drawn from pure log-normal statistics or from real data. Comparisons showed that \textsc{resolve} can outperform current imaging algorithms in these tasks. 

Furthermore, \textsc{resolve} is capable of producing an approximative uncertainty estimate for the inferred image through consistent propagation of measurement uncertainty. This is not possible with current imaging algorithms.

In addition to the inferred signal reconstruction, \textsc{resolve} also estimates the power spectrum of the signal, i.e. its two-point correlation structure. The power spectrum is used for the signal reconstruction, but can be regarded as a new scientific outcome by itself. For instance, it opens opportunities to study the statistical properties of magnetic fields that lead to observed synchrotron emission. At the same time it offers a unique tool to compare simulations of turbulent, magneto-ionic media in extended radio sources to observations.

It was shown that instead of using classical visibility weights directly, \textsc{resolve} chooses these internally, according to the ratio of reconstructed signal power to noise power. This is much in the spirit in which the robust weighting approach was originally conceived by Briggs \citep{BriggsThesis,BriggsPaper}.

It should be noted, however, that obtaining all results with extremely high accuracy, especially to produce the uncertainty map, can be more time consuming than traditional imaging methods because of the complicated numerical procedures necessarily involved to solve Eqs. (\ref{eq:one},\ref{eq:two}).

In this paper, only simulated data was analyzed and the fundamental principles underlying \textsc{resolve} were reviewed. To simplify the analysis, some typical complexities of radio interferometers have been omitted. However, the response operator $R$ (see Eq. \ref{response}), describing the act of observation, can easily be expanded to cover more effects, thereby adapting to the needs of the actual observational situation.

It is most straight-forward to include the effects of a primary beam, as long as it is known accurately for the instrument in question. Also a direction- or time-dependent point spread function can be included without any further fundamental complicacy, although computational complexity would be considerably higher. 


Furthermore, it should be highlighted that the inclusion of single dish data is almost readily possible. A radio interferometer is not sensitive to the largest scales of the sky brightness because it cannot measure at arbitrarily small $uv$-points, leaving a gap in the center of the $uv$-plane. This problem can in principle be overcome by combining the radio interferometric data with single dish observations on the same source. Always, when using CLEAN-derived imaging algorithms, there is a problem with the choice of the correct restoring beam, since it is not possible to just trivially use the point spread function of the radio interferometer for the combined data. There is no such problem with the imaging approach presented in this work.

The extension to multi-frequency synthesis (see Eq. \ref{Rmfs}) and polarization imaging is already being worked on and will be the subject of upcoming publications. 

Another future topic is the possible inclusion of calibration into the framework. A first step could be to include the calibrational errors into the error budget and use an approach similar to the extended critical filter \citep{ECF}, where the noise covariance is subject to the inference itself. In principle, calibration itself can be understood as a reconstruction problem for which the presented methods could be useful. In the long run, the distinction between \textit{calibration} and \textit{imaging} is somewhat artificial and should ideally be merged into one step of complete reconstruction \citep[see also][]{SmirnovI,SmirnovII}.

Finally, a future goal should be to extend the imaging algorithm \textsc{resolve} to a broader approach that can handle diffuse emission and point sources simultaneously \citep[see e.g.][for an example from photon count imaging]{d3po}. It could be worthwhile to think about merging the approaches of compressed sensing, where optimal imaging strategies for sparse signals are already known, with the presented Bayesian approach into which they could be included in form of a Laplacian prior.

\begin{acknowledgements}
We like to thank Niels Oppermann and Maksim Greiner for many helpful comments and discussions on statistics and numerics, Ashmeet Singh for extensive help with CASA, and Martin Reinecke and J\" org Knoche for computational support. H. Junklewitz thanks Rick Perley for the initial introduction to radio interferometry and its problems. Furthermore, we thank Oleg Smirnov, Annalisa Bonafede and Chris Hales for suggestions and discussions on the radio astronomical aspects of the work and especially Tracy C. Clarke for the same, and for kindly providing the data on the galaxy cluster Abell 2256. This work was partly conducted within the DFG Research Unit 1254 ``Magnetisation of Interstellar and Intergalactic Media'' and profited from the framework of the Magnetism Key Science Project of LOFAR.
\end{acknowledgements}

\bibliographystyle{aa}
\bibliography{references.bib}

\begin{thebibliography}{64}
\expandafter\ifx\csname natexlab\endcsname\relax\def\natexlab#1{#1}\fi

\bibitem[{{Aharonian} {et~al.}(2013){Aharonian}, {Arshakian}, {Allen},
  {Banerjee}, {Beck}, {Becker}, {Bomans}, {Breitschwerdt}, {Br{\"u}ggen},
  {Brunthaler}, {Catinella}, {Champion}, {Ciardi}, {Crocker}, {de Avillez},
  {Dettmar}, {Engels}, {En{\ss}lin}, {Enke}, {Fieseler}, {Gizon}, {Hackmann},
  {Hartmann}, {Henkel}, {Hoeft}, {Iapichino}, {Innes}, {James}, {Jasche},
  {Jones}, {Kagramanova}, {Kauffmann}, {Keane}, {Kerp}, {Kl{\"o}ckner},
  {Kokkotas}, {Kramer}, {Krause}, {Krause}, {Krupp}, {Kunz}, {L{\"a}mmerzahl},
  {Lee}, {List}, {Liu}, {Lobanov}, {Mann}, {Merloni}, {Middelberg}, {Niemeyer},
  {Noutsos}, {Perlick}, {Reich}, {Richter}, {Roy}, {Saintonge}, {Sch{\"a}fer},
  {Schaffner-Bielich}, {Schinnerer}, {Schleicher}, {Schneider}, {Schwarz},
  {Sedrakian}, {Sesana}, {Smol{\v c}i{\'c}}, {Solanki}, {Tuffs}, {Vetter},
  {Weber}, {Weller}, {Wex}, {Wucknitz}, \& {Zwaan}}]{FutureScienceReview}
{Aharonian}, F., {Arshakian}, T.~G., {Allen}, B., {et~al.} 2013, ArXiv e-prints

\bibitem[{{Beatty} {et~al.}(2005){Beatty}, {Nishimura}, \& {Pauly}}]{Beatty}
{Beatty}, P.~J., {Nishimura}, D.~J., \& {Pauly}, J.~M. 2005, IEEE Trans Med
  Imaging

\bibitem[{{Born} \& {Wolf}(1999)}]{Born}
{Born}, M. \& {Wolf}, E. 1999, {Principles of Optics}

\bibitem[{{Bracewell}(1965)}]{Bracewell}
{Bracewell}, R. 1965, {The Fourier Transform and its applications}

\bibitem[{{Briggs}(1995{\natexlab{a}})}]{BriggsPaper}
{Briggs}, D.~S. 1995{\natexlab{a}}, in Bulletin of the American Astronomical
  Society, Vol.~27, American Astronomical Society Meeting Abstracts, 112.02

\bibitem[{{Briggs}(1995{\natexlab{b}})}]{BriggsThesis}
{Briggs}, D.~S. 1995{\natexlab{b}}, PhD Thesis

\bibitem[{{Candes} {et~al.}(2006){Candes}, {Romberg}, \& {Tao}}]{CS1}
{Candes}, E.~J., {Romberg}, J.~K., \& {Tao}, T. 2006, Comm. Pure Appl. Math.

\bibitem[{{Carrillo} {et~al.}(2012){Carrillo}, {McEwen}, \& {Wiaux}}]{SARA}
{Carrillo}, R.~E., {McEwen}, J.~D., \& {Wiaux}, Y. 2012, \mnras, 426, 1223

\bibitem[{{Carrillo} {et~al.}(2013){Carrillo}, {McEwen}, \& {Wiaux}}]{PURIFY}
{Carrillo}, R.~E., {McEwen}, J.~D., \& {Wiaux}, Y. 2013, ArXiv e-prints

\bibitem[{{Caticha}(2008)}]{Caticha}
{Caticha}, A. 2008, ArXiv e-prints

\bibitem[{{Clark}(1980)}]{ClarkClean}
{Clark}, B.~G. 1980, \aap, 89, 377

\bibitem[{{Clarke} \& {Ensslin}(2006)}]{Tracy}
{Clarke}, T.~E. \& {Ensslin}, T.~A. 2006, \aj, 131, 2900

\bibitem[{{Cooley} \& {Tukey}(1965)}]{CooleyTukey}
{Cooley}, J.~W. \& {Tukey}, J.~W. 1965, Math. Comp., 19, 297

\bibitem[{{Cornwell}(2008)}]{MSClean}
{Cornwell}, T.~J. 2008, IEEE Journal of Selected Topics in Signal Processing,
  2, 793

\bibitem[{{Cornwell} \& {Evans}(1985)}]{MEM}
{Cornwell}, T.~J. \& {Evans}, K.~F. 1985, \aap, 143, 77

\bibitem[{{Cornwell} {et~al.}(2008){Cornwell}, {Golap}, \&
  {Bhatnagar}}]{WProjection}
{Cornwell}, T.~J., {Golap}, K., \& {Bhatnagar}, S. 2008, IEEE Journal of
  Selected Topics in Signal Processing, 2, 647

\bibitem[{{Donoho}(2006)}]{CS2}
{Donoho}, D.~L. 2006, IEEE Transactions of Information Theory

\bibitem[{{Eilek}(1989)}]{Eilek}
{Eilek}, J.~A. 1989, Bulletin of the American Physical Society, 34, 1286

\bibitem[{{En{\ss}lin}(2013)}]{IFT2}
{En{\ss}lin}, T. 2013, in American Institute of Physics Conference Series, Vol.
  1553, American Institute of Physics Conference Series, ed. U.~{von
  Toussaint}, 184--191

\bibitem[{{En{\ss}lin} \& {Frommert}(2011)}]{EnsFrom}
{En{\ss}lin}, T.~A. \& {Frommert}, M. 2011, \prd, 83, 105014

\bibitem[{{En{\ss}lin} {et~al.}(2009){En{\ss}lin}, {Frommert}, \&
  {Kitaura}}]{IFT}
{En{\ss}lin}, T.~A., {Frommert}, M., \& {Kitaura}, F.~S. 2009, \prd, 80, 105005

\bibitem[{{En{\ss}lin} \& {Weig}(2010)}]{EnsWeig}
{En{\ss}lin}, T.~A. \& {Weig}, C. 2010, \pre, 82, 051112

\bibitem[{{Finley} \& {Goss}(2000)}]{RadioSaga}
{Finley}, D.~G. \& {Goss}, W.~M., eds. 2000, {Radio interferometry : the saga
  and the science}

\bibitem[{{Garrett}(2012)}]{TelescopeReview}
{Garrett}, M.~A. 2012, in From Antikythera to the Square Kilometre Array:
  Lessons from the Ancients

\bibitem[{{Geman} \& {Geman}(1984)}]{GibbsSamp}
{Geman}, S. \& {Geman}, D. 1984, IEEE Transactions on Pattern Analysis and
  Machine Intelligence

\bibitem[{{Greiner}(2013)}]{MaksimMaster}
{Greiner}, M. 2013, The Galactic Free Electron Density: A Bayesian
  Reconstruction, Master Thesis

\bibitem[{{Gull} \& {Daniell}(1979)}]{GullMEM}
{Gull}, S.~F. \& {Daniell}, G.~J. 1979, in Astrophysics and Space Science
  Library, Vol.~76, IAU Colloq. 49: Image Formation from Coherence Functions in
  Astronomy, ed. C.~{van Schooneveld}, 219

\bibitem[{{Hastings}(1970)}]{MetroHast}
{Hastings}, W.~K. 1970, Biometrika, 57, 97

\bibitem[{{H{\"o}gbom}(1974)}]{1974A&AS...15..417H}
{H{\"o}gbom}, J.~A. 1974, \aaps, 15, 417

\bibitem[{{Huang}(1963)}]{Huang}
{Huang}, K. 1963, {Statistical Mechanics}

\bibitem[{{Jasche} {et~al.}(2010){Jasche}, {Kitaura}, {Wandelt}, \&
  {En{\ss}lin}}]{Jens}
{Jasche}, J., {Kitaura}, F.~S., {Wandelt}, B.~D., \& {En{\ss}lin}, T.~A. 2010,
  \mnras, 406, 60

\bibitem[{{Jaynes}(2003)}]{Jaynes}
{Jaynes}, E.~T. 2003, {Probability Theory: The Logic of Science}

\bibitem[{{Junklewitz} \& {En{\ss}lin}(2011)}]{Imprints}
{Junklewitz}, H. \& {En{\ss}lin}, T.~A. 2011, A\&A, 530, A88

\bibitem[{{Karakci} {et~al.}(2013){Karakci}, {Sutter}, {Zhang}, {Bunn},
  {Korotkov}, {Timbie}, {Tucker}, \& {Wandelt}}]{WandeltLine2}
{Karakci}, A., {Sutter}, P.~M., {Zhang}, L., {et~al.} 2013, \apjs, 204, 10

\bibitem[{{Lannes} {et~al.}(1997){Lannes}, {Anterrieu}, \&
  {Marechal}}]{WipeClean}
{Lannes}, A., {Anterrieu}, E., \& {Marechal}, P. 1997, \aaps, 123, 183

\bibitem[{{Lazarian} \& {Pogosyan}(2012)}]{Lazarian}
{Lazarian}, A. \& {Pogosyan}, D. 2012, \apj, 747, 5

\bibitem[{{Mood} {et~al.}(1974){Mood}, {Graybill}, \& {Duane}}]{stat}
{Mood}, A.~M., {Graybill}, F.~A., \& {Duane}, C.~B. 1974, Introduction to the
  theory of statistics

\bibitem[{{Neal}(1993)}]{HamiltonianSamp}
{Neal}, R.~M. 1993, Technical Report CRG-TR-93-1, Dept. of Computer Science,
  University of Toronto

\bibitem[{{Oppermann} {et~al.}(2011{\natexlab{a}}){Oppermann}, {Junklewitz},
  {Robbers}, \& {En{\ss}lin}}]{Probing}
{Oppermann}, N., {Junklewitz}, H., {Robbers}, G., \& {En{\ss}lin}, T.~A.
  2011{\natexlab{a}}, A\&A, 530, A89

\bibitem[{{Oppermann} {et~al.}(2011{\natexlab{b}}){Oppermann}, {Robbers}, \&
  {En{\ss}lin}}]{ECF}
{Oppermann}, N., {Robbers}, G., \& {En{\ss}lin}, T.~A. 2011{\natexlab{b}},
  \pre, 84, 041118

\bibitem[{{Oppermann} {et~al.}(2013){Oppermann}, {Selig}, {Bell}, \&
  {En{\ss}lin}}]{SmP}
{Oppermann}, N., {Selig}, M., {Bell}, M.~R., \& {En{\ss}lin}, T.~A. 2013, \pre,
  87, 032136

\bibitem[{{Peskin} \& {Schroeder}(1995)}]{PeskinSchroeder}
{Peskin}, M.~E. \& {Schroeder}, D.~V. 1995, {An Introduction to Quantum Field
  Theory} (Westview Press)

\bibitem[{{Rau} \& {Cornwell}(2011)}]{MSMFClean}
{Rau}, U. \& {Cornwell}, T.~J. 2011, \aap, 532, A71

\bibitem[{{Reid} \& {CASA Team}(2010)}]{CASA}
{Reid}, R.~I. \& {CASA Team}. 2010, in Bulletin of the American Astronomical
  Society, Vol.~42, American Astronomical Society Meeting Abstracts \#215,
  479.04

\bibitem[{{Rybicki} \& {Lightman}(1985)}]{RL}
{Rybicki}, G.~B. \& {Lightman}, A.~P. 1985, {Radiative processes in
  astrophysics.}

\bibitem[{{Ryle} \& {Hewish}(1960)}]{Ryle}
{Ryle}, M. \& {Hewish}, A. 1960, \mnras, 120, 220

\bibitem[{{Sault} \& {Oosterloo}(2007)}]{ImagingReview}
{Sault}, R.~J. \& {Oosterloo}, T.~A. 2007, ArXiv Astrophysics e-prints

\bibitem[{{Sault} \& {Wieringa}(1994)}]{MFClean}
{Sault}, R.~J. \& {Wieringa}, M.~H. 1994, \aaps, 108, 585

\bibitem[{{Schwab}(1984)}]{CSCLean}
{Schwab}, F.~R. 1984, \aj, 89, 1076

\bibitem[{{Selig} {et~al.}(2013){Selig}, {Bell}, {Junklewitz}, {Oppermann},
  {Reinecke}, {Greiner}, {Pachajoa}, \& {En{\ss}lin}}]{nifty}
{Selig}, M., {Bell}, M.~R., {Junklewitz}, H., {et~al.} 2013, \aap, 554, A26

\bibitem[{{Selig} \& {En{\ss}lin}(2013)}]{d3po}
{Selig}, M. \& {En{\ss}lin}, T. 2013, ArXiv e-prints

\bibitem[{{Selig} {et~al.}(2012){Selig}, {Oppermann}, \&
  {En{\ss}lin}}]{MatrixProbing}
{Selig}, M., {Oppermann}, N., \& {En{\ss}lin}, T.~A. 2012, \pre, 85, 021134

\bibitem[{{Skilling} {et~al.}(1979){Skilling}, {Strong}, \&
  {Bennett}}]{SkillingMEM}
{Skilling}, J., {Strong}, A.~W., \& {Bennett}, K. 1979, \mnras, 187, 145

\bibitem[{{Smirnov}(2011{\natexlab{a}})}]{SmirnovI}
{Smirnov}, O.~M. 2011{\natexlab{a}}, \aap, 527, A106

\bibitem[{{Smirnov}(2011{\natexlab{b}})}]{SmirnovII}
{Smirnov}, O.~M. 2011{\natexlab{b}}, \aap, 527, A107

\bibitem[{{Spangler}(1982)}]{Spangler1}
{Spangler}, S.~R. 1982, \apj, 261, 310

\bibitem[{{Spangler}(1983)}]{Spangler2}
{Spangler}, S.~R. 1983, \apjl, 271, L49

\bibitem[{{Sutter} {et~al.}(2012){Sutter}, {Wandelt}, \& {Malu}}]{WandeltLine1}
{Sutter}, P.~M., {Wandelt}, B.~D., \& {Malu}, S.~S. 2012, \apjs, 202, 9

\bibitem[{{Sutter} {et~al.}(2013){Sutter}, {Wandelt}, {McEwen}, {Bunn},
  {Karakci}, {Korotkov}, {Timbie}, {Tucker}, \& {Zhang}}]{SutterWandeltRadio}
{Sutter}, P.~M., {Wandelt}, B.~D., {McEwen}, J.~D., {et~al.} 2013, ArXiv
  e-prints

\bibitem[{{Taylor} {et~al.}(1999){Taylor}, {Carilli}, \& {Perley}}]{WhiteBook}
{Taylor}, G.~B., {Carilli}, C.~L., \& {Perley}, R.~A., eds. 1999, Astronomical
  Society of the Pacific Conference Series, Vol. 180, {Synthesis Imaging in
  Radio Astronomy II}

\bibitem[{{Thompson} {et~al.}(1986){Thompson}, {Moran}, \&
  {Swenson}}]{1986isra.book.....T}
{Thompson}, A.~R., {Moran}, J.~M., \& {Swenson}, G.~W. 1986, {Interferometry
  and synthesis in radio astronomy}

\bibitem[{{Transtrum} \& {Sethna}(2012)}]{TS12}
{Transtrum}, M.~K. \& {Sethna}, J.~P. 2012, ArXiv e-prints

\bibitem[{{Waelkens} {et~al.}(2009){Waelkens}, {Schekochihin}, \&
  {En{\ss}lin}}]{Waelkens}
{Waelkens}, A.~H., {Schekochihin}, A.~A., \& {En{\ss}lin}, T.~A. 2009, \mnras,
  398, 1970

\bibitem[{{Wiaux} {et~al.}(2009){Wiaux}, {Jacques}, {Puy}, {Scaife}, \&
  {Vandergheynst}}]{WiauxScaifePaper}
{Wiaux}, Y., {Jacques}, L., {Puy}, G., {Scaife}, A.~M.~M., \& {Vandergheynst},
  P. 2009, \mnras, 395, 1733

\end{thebibliography}

\appendix

\section{Derivation of \textsc{RESOLVE}}
\label{appendix1}

For a complete derivation of \textsc{resolve}, we first give some general remarks, and then divide the rest into two parts, where we derive a Maximum a Posteriori solution for the signal field and for its power spectrum, respectively.

From Sec.~\ref{SecTwo}, we recall the basic premises of the inference problem to be solved. We want to find the statistically optimal reconstruction of the total intensity signal $I$ given a data model

\begin{equation}
d = RI + n = R\mathrm{e}^{s} + n \label{appendix1: datamodel},
\end{equation}
under the assumptions that 

\begin{itemize}
 \item [-] $I$ follows log-normal statistics, such that $s=\log I$ follows Gaussian statistics,
 \item [-] the noise $n$ follows Gaussian statistics as well,
 \item [-] and $R$  models the linear response of a radio interferometer (see Eq. (\ref{response}) in Sec.~\ref{SecTwo}).
\end{itemize}
Under these assumptions the likelihood $\mathcal{P}(d|s)$ and the signal prior $\mathcal{P}(s)$ take the following form as was shown in (\ref{likeli2})

\begin{align}
\mathcal{P}(d|s) &= \mathcal{G}(d-R\mathrm{e}^{s},N) \notag\\
		 &= \frac{1}{\mathrm{det}(2 \pi N)^{1/2}} \ \mathrm{e}^{-1/2 \ \left((d-R\mathrm{e}^{s})^{\dagger} N^{-1} (d-R\mathrm{e}^{s})\right)}, \\
\mathcal{P}(s)   &= \mathcal{G}(s,S) \notag\\
		 &= \frac{1}{\mathrm{det}(2 \pi S)^{1/2}} \ \mathrm{e}^{-1/2 \ \left(s^{\dagger} S^{-1} s\right)}.
\end{align}
Then, the posterior of $s$ 

\begin{align}
\mathcal{P}(s|d) \propto \mathcal{G}(d-R\mathrm{e}^{s},N) \ \mathcal{G}(s,S) \label{appendix1: problem}
\end{align}
can become highly non-Gaussian due to the non-linearity introduced by (\ref{appendix1: datamodel}).

As a further complication, we have to assume \textit{a priori} that the signal covariance $S = \left<ss^{\dagger}\right>$ is unknown. Assuming statistical homogeneity and isotropy for the signal statistics,
we parameterize its power spectrum $P(k)$ as a decomposition into spectral parameters $p_i$ and positive projection operators $S^{(i)}$ onto a number of spectral bands such that the bands fill the complete Fourier domain
\begin{equation}
S = \sum_{i} p_i S^{(i)}. 
\end{equation}   

\textsc{resolve} consists of two inference steps to solve the main problem (\ref{datamodel}) iteratively for $s$ and all $p_i$. We fully describe both steps individually in the following subsections. 

  \subsection{Reconstruction of the signal field $s$}
  
  For the reconstruction of the signal field $s$, we assume the power spectrum parameters $p_i$ to be known from a previous inference step. This can formally be expressed by marginalizing over them while assuming a delta distribution for the known parameters $p^*$

  \begin{align}
  \mathcal{P}(s|d, p^*) &= \int \mathcal{D}p \ \mathcal{P}(s|d,p) \ \mathcal{P}(p|p^*) \notag\\
		        &= \int \mathcal{D}p \ \mathcal{P}(s|d,p) \ \delta(p - p^*).
  \end{align}
  
  For convenience, we rewrite our notation to work with the Hamiltonian $H(s,d)$ instead of the posterior $P(s|d)$ 

  \begin{equation}
  \mathcal{P}(s|d) := \frac{\mathrm{e}^{-H(d,s)}}{Z} \label{appendix1:Ham}
  \end{equation}
  with $Z := \mathcal{P}(d)$. This effectively expresses our problem in more familiar terms of statistical physics, while the Hamiltonian $H(s,d) = -\log\left(P(d|s)P(s)\right)$ still comprises all important signal-dependent terms and is usually easier to handle than the posterior.

  The Hamiltonian of problem (\ref{appendix1: problem}) reads
  \begin{align}
  H(s,d) &= -\log \left(\mathcal{G}(d-R\mathrm{e}^{s},N) \ \mathcal{G}(s,S)\right) \notag\\
       &= \frac{1}{2} \ s^\dagger S^{-1}_{p*} s + \frac{1}{2} (\mathrm{e}^s)^\dagger M \mathrm{e}^s - j^\dagger \mathrm{e}^s + H_{0} \label{appendix1: Hamfull}
  \end{align}
  where $j = R^\dagger N^{-1} d$, $M = R^\dagger N^{-1}R$ and $H_{0}$ summarizes all terms which are not dependent on the signal $s$. 

  Using the Gibbs free energy ansatz of \citet{EnsWeig}, \citet{SmP} have shown that it is possible to rederive the critical filter for this Hamiltonian. However, in practice, it is only solvable under the assumption of a diagonal $M$ in signal space. Otherwise we would be forced to explicitly compute arbitrary components of the very large matrix of size $n_s^{2}$, representing the operator$M$, which is computational infeasible. Unfortunately, for the response under consideration here (\ref{response}), with non-complete sampling of the Fourier plane in data space, $M$ will not be diagonal in general.

  Thus, we instead use the MAP principle to solve the inference problem for $s$. Maximizing the posterior readily translates to minimizing the Hamiltonian (\ref{appendix1:Ham}). If we take the derivative of the Hamiltonian (\ref{appendix1: Hamfull}) with respect to the signal field $s$ and set it to zero, we get

  \begin{align}
  \frac{\delta H(s)}{\delta s} = S^{-1}_{p^*} s + \mathrm{e}^s \cdot M \mathrm{e}^s - j \cdot \mathrm{e}^s = 0.
  \end{align}
  
  This is a high dimensional, non-linear equation, which can be solved numerically using an iterative optimization algorithm, in our case a steepest descent method. We call the solution of this equation $m = \mathrm{argmax}_s \mathcal{P}(s|d)$.

  The solution $m$ is an estimate for the Gaussian field $s$. To calculate a signal estimate $\hat{I}$ for the original log-normal signal $I = \mathrm{e}^s$, we just take the exponential of $m$

  \begin{equation}
  \hat{I} = \mathrm{e}^m. \label{estimate}
  \end{equation}

  \subsection{Uncertainty of the signal reconstruction}\label{UncertaintyTheory}

  A full statistical analysis involves accounting for the uncertainty of the signal estimate. For this, we use the information encoded in the second posterior moment (or covariance) $D = \langle (s-m) (s-m)^{\dagger}\rangle$ as a measure of the expected uncertainty of the signal reconstruction. Within the MAP approach, we approximate the inverse posterior covariance $D^{-1}$ with the second derivative of the Hamiltonian

  \begin{align}
  D^{-1} \approx -\frac{\delta^2 H(s)}{\delta s_x \ \delta s_y} \Large{|}_{s=m} &= \ S^{-1}_{p^* \ xy} +  \mathrm{e}^{s_x} M_{xy} \mathrm{e}^{s_y} \notag\\
					      & + \mathrm{e}^{s_y} \int \mathrm{d}z \ M_{xz} \ \mathrm{e}^{s_z} - j_x \cdot \mathrm{e}^{s_x} \ \delta_{xy}, \label{Dapprox}
  \end{align}
  which needs to be inverted numerically in practice. In this way, we effectively assume that the real signal posterior is approximated with a Gaussian $\mathcal{G}(m,D)$. Unfortunately, $D$ only approximates the posterior covariance of the Gaussian field $m$. We need to translate this into a posterior covariance for the full estimate $\hat{I} = \mathrm{e}^m$. 

  If the signal posterior \textit{were} exactly Gaussian, we could just assume our posterior estimate to be of exact log-normal statistics, solve for the mean and variance analytically and thus write

  \begin{align}
  \langle \mathrm{e}^{s_x} \rangle_{\mathcal{G}(m,D)} &= \mathrm{e}^{m_x + \frac{1}{2} D_{xx}} \label{lognormalmean}\\ 
  \langle \left(\mathrm{e}^{s_x}\right)^2 \rangle_{\mathcal{G}(m,D)} -  \langle \mathrm{e}^{s_x} \rangle_{\mathcal{G}(m,D)}^2 &= \mathrm{e}^{2m_x + D_{xx}} \left[\mathrm{e}^{D_{xx}} -1\right] \label{lognormalvar}
  \end{align}
  using the definitions for the mean and variance of a log-normal distribution \citep[see e.g.][]{stat}. But since the posterior is \textit{not} Gaussian in general, we cannot solve Eqs. (\ref{lognormalmean}, \ref{lognormalvar}) analytically. This was, in the first place, the reason why \textsc{resolve} uses the MAP approach (see Sec.~\ref{Sec:MainAlgo}). Nevertheless, since we effectively approximate the full posterior with a Gaussian $\mathcal{G}(m,D)$ when using Eq. (\ref{Dapprox}) as the posterior covariance, one might be tempted to just use Eqs.(\ref{lognormalmean}, \ref{lognormalvar}) anyhow.

  However, in practice, it turns out that within the MAP approach this procedure is prone to overestimating signal estimate and its uncertainty. This is because usually the maximum of a log-normal distribution lies \textit{above} its mean \citep[for details see][]{MaksimMaster}. We thus drop the extra terms of $D$ in the argument of the exponentials in Eqs.~(\ref{lognormalmean}, \ref{lognormalvar}), keep (\ref{estimate}), and write 

  \begin{equation}
  \hat{I_x} = \mathrm{e}^{m_x} \pm \sqrt{\mathrm{e}^{2m_x} \left[\mathrm{e}^{D_{xx}} -1\right]} \label{fullestimate}
  \end{equation}
  if we want to account for the uncertainty in the reconstruction.

  \subsection{Reconstruction of the power spectrum parameters $p$}

  In the second step of \textsc{resolve}, we assume to have a solution for $m$ and $D$ from the last iteration and estimate the unknown spectral parameters $p$ from the signal-marginalized probability of data and power spectrum $\mathcal{P}(p,d)$:
  
  \begin{align}
  \mathcal{P}(p,d) &= \int \mathcal{D}s \ \mathcal{P}(s,d|p) \ \mathcal{P}(p) \notag\\
	 &= \int \mathcal{D}s \ \mathcal{G}(d-R\mathrm{e}^s,N) \ \mathcal{G}(s,S_p) \ \mathcal{P}(p) \label{appendix1: sigmarpost}
  \end{align}

  This approach was first derived in \citet{SmP} for Gaussian signal fields. We closely follow their argument and show its approximate validity also for log-normal fields.

  In order to do this, we first need to define a prior for the power spectrum parameters $p$. In this, we follow \citet{EnsFrom}, \citet{EnsWeig} and \citet{SmP}, and choose independent inverse-gamma distributions for each spectral parameter $p_i$

  \begin{align}
  \mathcal{P}(p) &= \prod_i \mathcal{P}_\mathrm{IG}(p_i) \notag\\ \label{appendix1: inverse-gamma}
                 &= \prod_i \frac{1}{q_i \Gamma(\alpha_i - 1)} \left(\frac{p_i}{q_i}\right)^{-\alpha_i} \exp\left(-\frac{q_i}{P_i}\right).
  \end{align}
  where $\Gamma(\cdot)$ denotes the gamma function, $q_i$ defines an exponential cutoff in the prior for low values of $p_i$, and $\alpha_i$ is the slope of the power-law decay for large values of $p_i$. In principle, by tuning these parameters, the prior can be adapted according to the a priori knowledge about the power spectrum. Usually, we use the limits of $q_k \rightarrow 0$ and $\alpha_k \rightarrow 1$ for all k. This turns the inverse-gamma prior into Jeffreys prior \citep{Jaynes}, which is flat on a logarithmic scale. In some tests though, we have allowed for non-unity $\alpha_k$ parameters to suppress unmeasured Fourier modes. 

  During the reconstruction of the power spectrum, we additionally introduce a smoothness prior as developed by \citet{SmP} to punish most probably unphysical and numerically unwanted random fluctuations in the power spectrum. In that prescription, the inverse-gamma prior (\ref{appendix1: inverse-gamma}) is augmented with a probability distribution that enforces smoothness of the power spectrum

  \begin{equation}
    \mathcal{P}(p) = \mathcal{P}_\mathrm{sm}(p) \prod_k \mathcal{P}_\mathrm{IG}(p_k).
  \end{equation}
  The  spectral smoothness prior can be written as a Gaussian distribution in $\tau = \log p$:
  
  \begin{align}
    \label{appendix1: smoothnesspriorwithT}
    \mathcal{P}_\mathrm{sm}(p) &\propto \exp \left( -\frac{1}{2 \sigma_p^2} \int \!\! \mathrm{d}{\left(\log k\right)}\, \left(\frac{\partial^2 \log p_k}{\partial \left(\log k\right)^2}\right)^2 \right) \notag\\
    			       &\propto \exp \left( -\frac{1}{2} \tau^\dagger T \tau \right),
  \end{align}
  where the differential operator $T$ includes the second derivative of $\tau = \log p$ and a scaling constant $\sigma_p^2$ that determines how strict the smoothness should be enforced. This particular form of the prior favors smooth power-law spectra. For all details we refer to \citep{SmP}. 

  As was shown there, the corresponding inverse-gamma prior for the $\tau$ parameters can easily be derived from the conservation of probability under transformations

  \begin{align}
    \mathcal{P}(\tau) &= \mathcal{P}(p)\, \left|\frac{\mathrm{d}p}{\mathrm{d}\tau}\right|\notag\\
    &= \prod_i \frac{q_i^{\alpha_i - 1}}{\Gamma(\alpha_i - 1)} \mathrm{e}^{-\left[\left(\alpha_i - 1\right) \tau_i + q_i \mathrm{e}^{-\tau_i}\right]}.
  \end{align}
  
  With this prior, we can calculate the signal-marginalized joint probability (\ref{appendix1: sigmarpost}) if we apply one crucial approximation. Since $\mathcal{P}(s,d|\tau)$ in (\ref{appendix1: sigmarpost}) is non-Gaussian due to the high non-linearity of the $\mathrm{e}^{(d-R\mathrm{e}^s)}$ - terms, we cannot just move on analytically. We instead use a saddle point method and approximate the argument of the exponential occurring in $\mathcal{P}(s,d|\tau)$, which can be written as $\mathrm{e}^{-H(s,d)}$ using (\ref{appendix1:Ham}). To perform the saddle point approximation, we replace  $H(s,d)$ with its Taylor expansion up to second order around the maximum of the Posterior $m$, derived in the previous iteration of the signal reconstruction: 

  \begin{align}
  \mathrm{e}^{-H(s,d)} &\propto \mathrm{e}^{\left(-\frac{1}{2} (d - R\mathrm{e}^s)^{\dagger} N^{-1} (d - R\mathrm{e}^s) - \frac{1}{2} s^{\dagger} S_{\tau}^{-1} s\right)} \notag\\
		       &\approx \mathrm{e}^{\left(H(m) + \frac{1}{2} (s - m)^{\dagger} D(m)^{-1} (s-m)\right)} \label{appendix1: approx}
  \end{align}
  This effectively approximates the non-Gaussian signal posterior $\mathcal{P}(s,d|\tau)$ with a Gaussian with mean $m$ and covariance $D$.
  We note that this procedure is similar to a mean field approximation in statistical physics \citep{Huang}.

  With this approximation, we can solve the marginalization integral in (\ref{appendix1: sigmarpost}) and calculate $\mathcal{P}(\tau,d)$, or alternatively the Hamiltonian
 
  \begin{align}
    H(d,\tau) &= -\log \mathcal{P}(d,\tau)\nonumber\\
    &= -\log \int \mathcal{D}s\, \mathcal{G}(d-R\mathrm{e}^s,N)\, \mathcal{G}(s,S)\, \mathcal{P}(p)\notag\\
    &\approx \frac{1}{2} \mathrm{tr} \left(\log S_{\tau} \right) - \frac{1}{2} \mathrm{tr} \left(\log D_{\tau} \right) + H(m,\tau) \notag\\
    \label{appendix1: H-smooth}
    &~~~ + \sum_i \left(\left(\alpha_i - 1\right) \tau_i + q_i \mathrm{e}^{-\tau_i}\right) + \frac{1}{2} \tau^{\dagger} T \tau \notag\\
    &~~~ + H_0 
  \end{align}
  where we have used the matrix theorem $\log |S| = \mathrm{tr} \left(\log S \right)$, and have collected all terms not depending on $\tau$ into a constant $H_0$.

  Taking the derivative of (\ref{appendix1: H-smooth}) with respect to one parameter $\tau_i$ and replacing $p_i = \mathrm{e}^{\tau_i}$, we find

  \begin{align}
  p_i = \frac{q_i + \frac{1}{2} \mathrm{tr} \left((mm^{\dagger} + D)S^{(i)}\right)}{\alpha_i - 1 + \frac{\varrho_i}{2} + (T\log{p})_i}. \label{appendix1: CF}
  \end{align}

  With this equation we can update the power spectrum parameters for each iteration using the current $m$ and $D$.

  This is in perfect accordance with previous findings \citep{EnsFrom,EnsWeig,SmP} and shows effectively that we can re-discover the critical filter for a pure MAP approach if we accept the approximation (\ref{appendix1: approx}) as valid.

\section{Implementation of \textsc{RESOLVE}} 
\label{appendix2}

\subsection{General implementation}
We have implemented \textsc{resolve} in \textsc{Python}, where crucial parts have been translated into more efficient \textsc{C} code using \textsc {Cython}\footnote{See http://docs.cython.org/.} . The actual implementation of the algorithm makes heavy use of the versatile inference library \textsc{NIFTy} \citep{nifty}. 

To perform the gridding and degridding operations needed in radio astronomical applications, we use the generalized Fast Fourier transformations package \textsc{gfft} \footnote{See https://github.com/mrbell/gfft}. The grid convolution is performed using a Kaiser-Bessel kernel following \citep{Beatty}.

For numerical optimization, we use a self-written steepest descent solver and in some cases the conjugate gradient routine provided by the \textsc{SciPy} package. 

The algorithm is controlled by a number of numerical procedures and parameters, governing the grade of convergence and the degree of accuracy. Apart from standard parameters, such as the maximum number of iterations or the accuracy of the steepest descent, the most important are:

\begin{itemize}
 \item [-] Different starting guesses for $s$ and $p$ might have strong impact on the performance or the solution of \textsc{resolve}. In non-linear optimization, there is, for instance, always the danger to only converge to a local minimum. Experience showed that in most cases, it is optimal to use constant fields and simple generic power spectra as starting guesses to prevent any biases. But other options are available, e.g. a CLEAN or a dirty map, and/or their respective empirical power spectra, in some cases allowing for an improvement in computation time.
 \item [-] To calculate $D$ for (\ref{appendix1: CF}), we have to numerically invert $D^{-1}$ and statistically probe the needed matrix entries \citep{MatrixProbing} using an implicit representation of the operator as a coded function. For this, we employ a conjugate gradient routine whose convergence and accuracy parameters must be set. This numerical inversion is usually the most serious bottleneck in computation time (see Sec.~\ref{algoeff}). Especially calculating $D$ for an estimate of the signal uncertainty can be a time consuming task, depending on the accuracy needed.
 \item [-] For observations with rather poor $uv$-coverage, problems might occur with the inversion of the operator $D$, which sometimes tends to be numerically non-positive definite during early iterations. In that case, we have implemented a solution where a diagonal matrix with a user-defined positive constant $M_0$ gets added to $D^{-1}$ to ensure positive-definiteness. While the solution is slowly converging over the global iterations, $M_0$ is constantly decreased. This is a standard approach in numerical optimization, see for instance \citet{TS12}.
 \item [-] For large data sets, it is sometimes of high advantage to bin the power spectrum instead of mapping it over all possibly allowed modes set by the user defined image size. Otherwise, the calculation might take prohibitively long.   
\end{itemize}

\subsection{Analysis of algorithmic efficiency}\label{algoeff}

As visualized in Fig.~\ref{fig:flow chart}, \textsc{resolve} mainly consists of two parts, a signal estimator, and a power spectrum estimator. They are iterated $N_{\mathrm{global}}$ times, until convergence is achieved, while both the maximum number of iterations and the exact convergence criteria can be set be the user. The signal estimator utilizes a steepest descent algorithm to solve Eq.~(\ref{eq:one}), which needs $N_{\mathrm{sd}}$ internal iterations. The power spectrum is estimated with Eq.~(\ref{eq:three}), where the trace of the inverse operator given by Eq.~(\ref{eq:two}) needs to be calculated. Since the operator is only given implicitly, its diagonal entries need to be probed $N_{\mathrm{pr}}$ - times using random vectors \citep{MatrixProbing}, where, for each probe, the operator equation (\ref{eq:two}) has to be inverted using a conjugate gradient algorithm.  

The steepest descent iterations are dominated by the operations needed to calculate $M$ (see Eq.~(\ref{eq:one})), which involves the response operator $R$ with a FFT and a subsequent Gridding operation. Therefore, its computational cost goes roughly with $N_{\mathrm{sd}}\left(O(n_d) + O(n_s \log(n_s))\right)$, where $n_d$ is the total number of visibilities, and $n_s$ the number of pixels in image space. 

The conjugate gradient is dominated by the need to compute the same operation, only, at least some fraction of $n_s$ times, and for each probe individually. Usually a maximum of $\sqrt{n_s}$  iterations of the conjugate gradient are performed. This leads to a total computational cost of roughly $N_{\mathrm{pr}}\left(O(\sqrt{n_s} n_d) + O(\sqrt{n_s} n_s \log(n_s))\right)$.

A realistic assessment of the asymptotic overall algorithmic efficiency is complicated, because all of the iteration numbers, $N_{\mathrm{global}}$, $N_{\mathrm{sd}}$, and $N_{\mathrm{pr}}$ can in principle vary strongly from case to case. Although $N_{\mathrm{sd}}$ usually will be larger than $N_{\mathrm{pr}}$\footnote{At least empirically taken from the simulations, the number of probes can be kept well below a couple of hundreds.}, the conjugate gradient term will likely dominate the algorithmic costs. In realistic applications, $n_d$ will usually be larger than $n_s$, because, for modern instrument data sets, the number of visibilities can reach the millions. In that case, the algorithmic efficiency probably tends to $N_{\mathrm{global}} N_{\mathrm{pr}} O(\sqrt{n_s} n_d)$.

In addition, this analysis shows that calculating an estimate for the uncertainty of the signal reconstruction is very costly. To accurately compute the diagonal of $D$, a large number of probes is needed so that $N_{\mathrm{pr}}$ can easily exceed the thousands.

On our development machine, with up to 8 used CPUs and a maximum of 64GB working memory, the non-optimized code produced the results presented in Sec.~\ref{mainres} in roughly a couple of hours for the low noise case, and a couple of days for the high noise case. For the relatively small size of the simulated VLA snapshot data sets, we never used more than a few percent of the memory but this would most likely change for larger data sets.

\section{A signal inference view on visibility weighting}
\label{appendix3}

In radio astronomy, the imaging step of aperture synthesis is usually combined with a weighting scheme that is included in the Fourier inversion of the visibilities. Essentially, the term $W$ in (\ref{invprob}), defining the dirty image $I^{\mathrm{D}}$, can be expanded to hold more factors than the mere sampling function

\begin{align}
I^{\mathrm{D}} =  \mathcal{F}^{-1}(T \cdot B \cdot w \cdot S \cdot \mathcal{F}I) 
\end{align}
with $W = T \cdot B \cdot w \cdot S$, where $T$ is a possible tapering of outer visibilities, $B$ is a user-defined baseline weighting, $w$ are the statistical noise weights obtained from an analysis off the thermal noise, and $S$ is the sampling function.

Historically, mainly two weighting schemes have been employed. \textit{Natural} weighting just multiplies every visibility point with the inverse thermal noise variance for the particular baseline and is therefore a simple, noise-dependent down-weighting mechanism. \textit{Uniform} weighting ensures that the weight per gridded visibility cell is constant and, hence, effectively gives higher weight to outer baselines, where usually less visibility points are found in a grid cell.

In a seminal work \citep{BriggsThesis}, Briggs has shown that natural weighting can be obtained under the constraint that the sample variance of the image should be minimized. In contrast, uniform weighting can be shown to reduce sidelobe levels, but actually downgrades sensitivity at the same time. 

In the same work, a new weighting scheme was devised that interpolates between these two extremes, called \textit{robust} weighting. The robust weights are determined as

\begin{align}
W(k) \propto \frac{1}{1+\sigma^2(k)/s^2_{\mathrm{p}}(k)} \label{robust weighting}
\end{align}
where $\sigma^2$ is the thermal noise variance, and $s^2_{\mathrm{p}}$ is some parameter that originally was derived having in mind some measure of the source power at the given visibility \citep{BriggsThesis}. In practice, $s^2_{\mathrm{p}}$ is usually adjusted by hand to meet the needs of the astronomer for having a tradeoff between sensitivity and resolution.

This form of weighting can be explained within the presented Bayesian framework, and, furthermore, we will show that an algorithm like \textsc{resolve} \textit{automatically} chooses the optimal weighting parameters according to the ratio of estimated noise and signal power.

For this, we consider the negative logarithm of the posterior (\ref{posterior}), i.e. the Hamiltonian of our inference problem (see Eq.~(\ref{appendix1: problem}) in App.~\ref{appendix1} for details)

\begin{align}
H(s,d) = \frac{1}{2} \ s^\dagger S^{-1} s + \frac{1}{2} (\mathrm{e}^s)^\dagger M \mathrm{e}^s - j^\dagger \mathrm{e}^s + H_{0} \label{weightHam}.
\end{align}

We can expand the exponents in a Taylor series and separate the quadratic from the higher orders in $s$ as we have done in (\ref{FullIFT}):

\begin{align}
H(s,d) &=  \frac{1}{2} \ s^{\dagger}\left(S^{-1}+M\right)s \ - \ s^{\dagger}j + H_0 \notag\\
       &+ \ \sum\limits_{k=3}^{\infty} \frac{1}{k!} \Lambda(M,j)^{k}_{x_{1} \cdots x_{k}} s_{x_{1}} \cdots s_{x_{k}}.
\end{align}

If we now apply the MAP principle and set the derivative with respect to $s$ to zero, we find 

\begin{align}
\left(S^{-1}+M\right)s \ -j \ + \ \Delta(M,j,s) = 0
\end{align}
where we have defined $\Delta(M,j,s)=\frac{\delta}{\delta s} \ \sum\limits_{k=3}^{\infty} \frac{1}{k!} \Lambda(M,j)^{k}_{x_{1} \cdots x_{k}} s_{x_{1}} \cdots s_{x_{k}}$. We can partly solve this equation for s:

\begin{align}
s = \left(S^{-1}+M\right)^{-1}j \ - \left(S^{-1}+M\right)^{-1} \Delta(M,j,s). \label{WF-Full-equation}
\end{align}

The first term is the analytic solution to the quadratic part of the full log-normal Hamiltonian. It was shown to be equivalent to a Wiener Filter applied to the data $d$ \citep{IFT}, which would be the optimal solution for a purely Gaussian signal field.

Using (\ref{PS}) for the covariance matrices $S$ and $N$ and $j = R^{\dagger}N^{-1}d$, we can write the Wiener Filter operator in (\ref{WF-Full-equation}), $F = \left(S^{-1}+M\right)^{-1} R^{\dagger}N^{-1}$, in Fourier space:

\begin{align}
F(k) = \frac{1}{1+P^{g}_{n}(k)/P_{s}(k)}
\end{align}
where $P^{g}_{n} = G_{ku} P_{n}(u)$ is the noise power spectrum on the regular grid, defined by the gridding operator $G$ from (\ref{gridresponse}).

This has the exact same form as the definition of the robust weights (\ref{robust weighting}), and even the original premise is fulfilled that the factor $s_{p}^2$ in (\ref{robust weighting}) should be connected to the source power. The great difference is that the Wiener Filter automatically weights each mode in Fourier space differently, given that the signal power spectrum $P_{s}(k)$ is known. 

We conclude that the classical robust weighting can be theoretically understood as the optimal solution to a signal reconstruction problem of a Gaussian signal field, equivalent to a Wiener Filter operation. In fact, this similarity between the robust weights and Wiener Filtering was already mentioned by Briggs himself \citep{BriggsThesis}, although in that work, no clear explanation of the connection was given.  

In common practice, of course, the weights are set manually, as only the knowledge of the signal power spectrum would allow for an automatic assignment. Since \textsc{resolve} reconstructs this power spectrum, it does implicitly assign these weights. Of course, \textsc{resolve} solves (\ref{WF-Full-equation}) iteratively, and only the converged solution will give optimal weights for the log-normal inference problem. No simple and direct equivalence can be given between these effective weights and robust weighting. It is even unclear how to write them down explicitly since the sum in $\Delta(M,j,s)$ in principle extends infinitely.

\end{document}